\documentstyle[emulateapj]{article}

\lefthead{Alonso-Herrero et al.}
\righthead{The Nature of LINERs}

\begin{document}
\title{The Nature of LINERs\footnote{Observations
reported in this paper were obtained with the
Multiple Mirror Telescope, which is operated jointly by the Smithsonian
Astrophysical Observatory and the University of Arizona.}}

\author{Almudena Alonso-Herrero, Marcia J. Rieke, George H. Rieke}
\affil{Steward Observatory, The University of Arizona, Tucson, AZ 85721}
\and
\author{Joseph C. Shields}
\affil{Department of Physics and Astronomy, Ohio University, Athens,
OH 45701-2979}

\begin{abstract}
We present $J$-band ($1.15-1.35\,\mu$m) spectroscopy of a sample of nine
galaxies showing some degree of LINER activity (classical
LINERs, weak-[O\,{\sc i}] LINERs and transition objects),
together with $H$-band  spectroscopy
for some of them. A careful subtraction of the stellar continuum
allows us to obtain reliable [Fe\,{\sc ii}]$1.2567\,\mu$m/Pa$\beta$
line ratios. We conclude that different types of LINERs (i.e.,
photoionized by a stellar continuum or
by an AGN) cannot be easily distinguished based solely on
the [Fe\,{\sc ii}]$1.2567\,\mu$m/Pa$\beta$ line ratio.

The emission line properties of many LINERs
can be explained in terms of an aging starburst. The optical
line ratios of these LINERs are reproduced by a model with a metal-rich
H\,{\sc ii} region component photoionized with a single stellar temperature $T_* = 38,000\,$K, plus a supernova remnant (SNR) component.
The [Fe\,{\sc ii}] line is predominantly excited by
shocks produced by SNRs in starbursts and starburst-dominated LINERs, while Pa$\beta$ tracks H\,{\sc ii} regions ionized by massive young stars. The contribution from SNRs to the overall emission line spectrum is constrained by the [Fe\,{\sc ii}]$1.2567\,\mu$m/Pa$\beta$ line ratio. Although our models for aging starbursts are constrained only by these infrared lines, they consistently explain the optical spectra of the galaxies also. 

The LINER-starburst connection is tested by predicting the time dependence
of the ratio of the ionizing luminosity  ($L_{\rm ion}$)
to the supernova rate (SNr), $L_{\rm ion}$/(SNr). We 
predict the relative number of starbursts to
starburst-dominated LINERs (aging starbursts) and show that it is in approximate agreement with survey findings for nearby galaxies. 

\end{abstract}


\section{INTRODUCTION}
\label{sec:intr}

Low ionization nuclear emission line regions (LINERs, Heckman 1980)
are probably the most common type of galaxy nuclear activity;
1/3 of all spiral galaxies with $B<12.5$ show LINER spectra
(Ho, Filippenko, \& Sargent 1993, 1997b).
LINERs, however, appear to be a heterogeneous class of objects (Heckman
1986). The detection of a broad H$\alpha$ component in some ($\sim 24\%$)
LINERs (Ho et al. 1993, 1997b), the presence
of unresolved UV sources in a few objects (Maoz et al. 1995)
and the continuity of X-ray properties from the luminous active Seyferts
to low-luminosity active galaxies (Koratkar et al. 1995)
suggest that in some LINERs photoionization
by a non-stellar continuum (AGN) source is
the dominant excitation process. However LINER spectra can also be modeled
by photoionization by hot stars in a dense medium (Shields 1992;
Filippenko \& Terlevich 1992). The similarity of some LINER spectra to those
of supernova remnants initially suggested that the optical lines of LINERs
could be excited by shock heated gas (Heckman 1980) caused by supernovae or
from AGN-associated processes.

Engelbracht et al. (1998) show in detail how the LINER
characteristics of the weak-[O\,{\sc i}] LINER/H\,{\sc ii} galaxy NGC~253 arise
unavoidably from an old nuclear starburst, as its hot stars fade and a high
supernova rate persists. Other authors have proposed less specific connections
between LINER emission-line characteristics and late phase starbursts. For
example, Larkin et al. (1998) suggest that a number of classical LINERs have
Pa$\beta$ absorption stronger than would be expected for a pure old stellar
population, indicating the presence of a significant number of hotter stars.
Maoz et al. (1998) find that UV spectra of LINERs can show stellar
absorption lines, suggesting that the emission of some LINERs with compact UV
sources could be unrelated to the presence of a low luminosity AGN.

\begin{deluxetable}{lcccccc}
\tablefontsize{\footnotesize}
\tablecaption{The sample of LINERs.}
\tablehead{\colhead{} & \colhead{$v_{\rm hel}$} & \colhead{} &
\colhead{$t_{\rm exp}$} & \colhead{Nuclear} & \colhead{Reference} &
\colhead{Reference} \\
\colhead{Galaxy}  & \colhead{(km s$^{-1}$)} & \colhead{Run} &
\colhead{(s)} & \colhead{Type} & \colhead{optical} & \colhead{infrared}\\
\colhead{(1)}    & \colhead{(2)}& \colhead{(3)} & \colhead{(4)} &
\colhead{(5)} & \colhead{(6)} & \colhead{(7)}}
\startdata
NGC~2639 & 3336 & Dec 1996 & 2880 & LINER &  1,2,7 & a \nl
NGC~3031 (M81) & -34  & Dec 1996 & 1440 & LINER &  1,2 & b \nl
NGC~3367& 3037 & Dec 1996 & 1920 & weak-LINER & 1,3 & a \nl
NGC~3504& 1539 & Dec 1996 & 1440 & weak-LINER & 1,2 & a \nl
NGC~3998 & 1040 & Dec 1996 & 2160 & LINER & 1,2,7 & a \nl
NGC~4569 (M90) & -235 & Dec 1996 & 2880 & weak-LINER & 1,4,7 & c \nl
NGC~4579 (M58) & 1519 & Apr 1997 & 2160 & LINER & 1,5,7 & a \nl
NGC~5953 & 1965 & Apr 1997 & 2160 & LINER/Sy2/SB & 6 & a \nl
NGC~7743 & 1710 & Dec 1996 & 1920 & LINER/Sy2 & 1,2 & a \nl
\hline
NGC~3379 & 920 & Mar 1997 & 2160 & E (template) & \nodata & \nodata \\
\enddata
\medskip
\tablecomments{Col.~(1) Galaxy. Col.~(2) Heliocentric velocity from NED.
Col.~(3) Observing run for the near-IR spectroscopy. Col.~(4)
Integration time for the $J$-band spectroscopy.
Integration times for $H$-band spectra: NGC~3367 5760\,s and
4320\,s,  NGC~3504 4320\,s (intermediate resolution) and
1680\,s (high resolution), NGC~4569 1440\,s and NGC~7743 1920\,s.
Col.~(5) Nuclear type, see Appendix for more
details on the classification. Cols.(6)
References for the optical line ratios:
1. Ho et al. (1997a) 2. Ho et al. (1993) 3. V\'eron-Cetty
\& V\'eron (1986) and V\'eron et al. (1997) 4. Stauffer (1982)
5. Gonz\'alez-Delgado \& P\'erez (1996b) 6. Veilleux et al. (1995). 7.
Keel (1983). Col.~(7) References for the infrared aperture photometry:
a. This work b. Forbes et al. (1992) c. Devereux et al. (1987)}
\end{deluxetable}

Infrared and extreme emission line galaxies appear to be undergoing starbursts
with a duration of $\sim$ 10 million years (Rieke et al. 1993; 
Genzel et al. 1995; Engelbracht et al. 1996, 1998; B\"oker, 
F\"orster-Schreiber, \& Genzel 1997; Vanzi, Alonso-Herrero, \&
Rieke  1998).  As starbursts age, they should
fade to become Balmer absorption galaxies. However,
the intermediate age starbursts ($\sim 20 - 40$ million years old) have not
been found in significant numbers. The modeling by Engelbracht et al. (1998)
suggests that many of these galaxies have been masquerading as
LINERs. The goal of this study is to test this connection by making
use of near-IR spectroscopy and additional modeling of a sample of nearby
LINERs. In Section~2 we describe the
sample and the observations. In Section~3, we derive near-infrared and optical
line ratios and show that in many cases they agree with expectations for aging
starbursts. In Section~4 we show that the
LINER-starburst connection predicts a plausible number of LINERs to agree with
observation. Our conclusions are presented in Section~5.

\section{OBSERVATIONS}
\subsection{The sample}

We present in this study near-infrared spectroscopy of a sample of galaxies
showing some degree of LINER activity. Due to the heterogeneous nature of
the LINER class we have adopted the following initial criteria for the
classification. We selected candidates
from Ho et al. (1993) and Ho et al. (1995); for the latter reference, the
LINER classification was confirmed with optical line ratios found elsewhere in
the literature (references given in Table~1). We classified a galaxy as a {\it
pure} (or classical) LINER only if it fully satisfies Heckman's (1980)
definition ([O\,{\sc ii}]$\lambda 3727 > $[O\,{\sc iii}]$\lambda$5007 and
[O\,{\sc i}]$\lambda 6300 > 0.3\times$[O\,{\sc iii}]$\lambda$5007).
Weak-[O\,{\sc i}]6300 LINERs were taken to have the line ratio [O\,{\sc
i}]$\lambda$6300/H$\alpha < 1/6$ (Filippenko \& Terlevich 1992). Other
galaxies can be classified as transition objects because they show
a composite H\,{\sc ii}/LINER spectra. In addition, the activity
classification seems to be aperture dependent for some objects in our
sample. In the Appendix we give more details on the classification of
each galaxy, since in some cases the determination of the LINER class is
not straightforward.

For the initial sample we used the H$\alpha$ fluxes from the literature
to determine what galaxies might have a detectable level of Pa$\beta$
emission at our resolution. Our sample is therefore biased toward
relatively bright emission line galaxies. Such a bias is inevitable for
an infrared spectroscopic sample,
since the Pa$\beta$ line is intrinsically an order of magnitude weaker than
H$\beta$ and a normal galaxy continuum at constant spectral resolution is an
order of magnitude brighter at Pa$\beta$ than at H$\beta$, so the emission
line spectrum is detectable in the infrared only for strong-lined galaxies.

For the purposes of this study we will also include the near-infrared
spectroscopic data of the LINER prototype NGC~1052 (Alonso-Herrero et al.
1997), the weak-[O\,{\sc i}] LINER NGC~253 (Engelbracht et al. 1998),
the classical LINER/Wolf-Rayet galaxy NGC~6764 (Calzetti 1997), and
the sample of LINERs presented in Larkin et al. (1998).

\begin{figure*}
\figurenum{1}
\plotfiddle{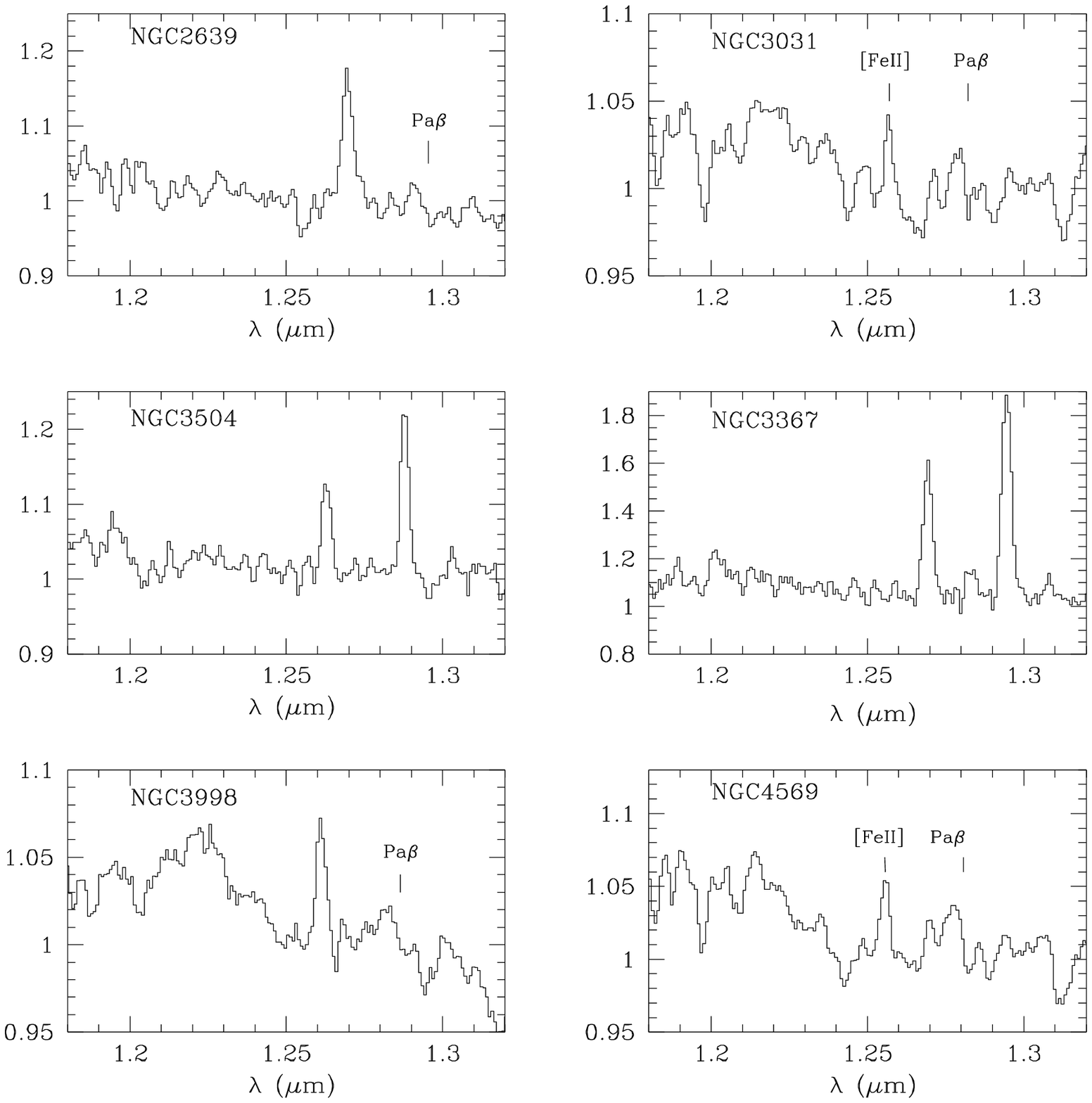}{425pt}{0}{60}{60}{-200}{0}
\plotfiddle{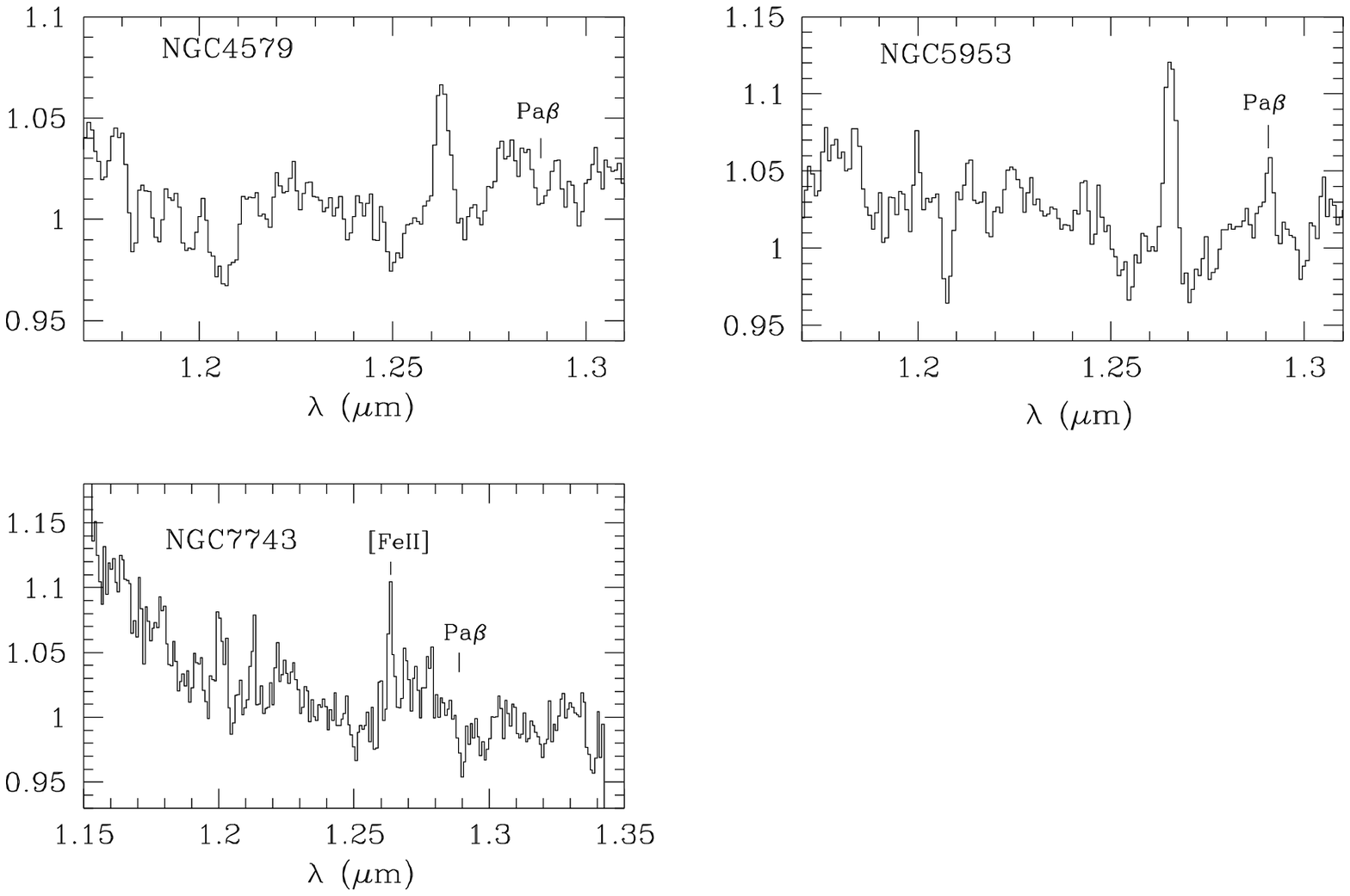}{425pt}{0}{60}{60}{-200}{100}
\vspace{-10cm}
\caption{Observed $J$-band spectra of the LINERs normalized to unity.
The extraction aperture is $1.2\arcsec \times 2\arcsec$ in all cases.
The [Fe\,{\sc ii}]$1.257\,\mu$m is clearly
detected in all spectra.}
\end{figure*}

\subsection{Near-Infrared Spectroscopy}
We obtained near-infrared long-slit spectra in the $J$-band
($1.15-1.35\,\mu$m) of nine LINERs at the Multiple Mirror
Telescope (MMT) with the infrared spectrometer FSPEC (Williams et \
al. 1993) during
two observing runs in 1996 December and 1997 April. For both observing
runs we used a slit of $1\farcs2 \times 30\arcsec$ with
pixel size 0\farcs4 pixel$^{-1}$
and a  300 groove mm$^{-1}$ grating which provides a resolution of
$\lambda/\Delta\lambda \simeq 460$ at
$1.25\,\mu$m. In addition to the spectra of the LINERs, a $J$-band spectrum of
the elliptical galaxy NGC~3379, to be used as a template, was obtained at the
Steward Observatory 61-Inch telescope on 1997 March 26, with pixel size
1\farcs8 pixel$^{-1}$ and a slit two pixels wide.

Observations were obtained for each galaxy at three or four positions along
the slit, integrating for 2 or 4 minutes at each position. This pattern was
repeated until the desired signal to noise was achieved (at least 40, if
possible). Approximately solar-type stars were
observed in a similar fashion, interspersed with
the galaxy observations, and selected  to be at similar air
masses. The data reduction process involves dark current
subtraction, flat-fielding and sky subtraction. The correction for
atmospheric transmission is performed by dividing the galaxy spectrum
by the adjacent spectrum of the standard star.
The resulting spectrum is multiplied by a solar spectrum to
correct for the standard star absorption features (as described by Maiolino,
Rieke, \& Rieke 1996). The wavelength
calibration was performed by making use of OH sky lines from the list
in Oliva \& Origlia (1992).

Additional spectra at resolution $\lambda/\Delta\lambda \simeq 750$
in the $H$-band ($1.65\,\mu$m) were obtained for NGC~3367, NGC~3504 and NGC~4569 on the 1996 Dec run. A  high resolution
($\lambda/\Delta\lambda \simeq 1500$) spectrum in the
$H$-band ($\lambda = 1.68\,\mu$m)
of NGC~3504 was also obtained on the night of 1996 June 1 which will be
combined with the $\lambda = 1.63\,\mu$m spectra presented in
Alonso-Herrero et al. (1997). Similar observational and reduction
procedures were used throughout.

\begin{figure*}
\figurenum{2}
\plotfiddle{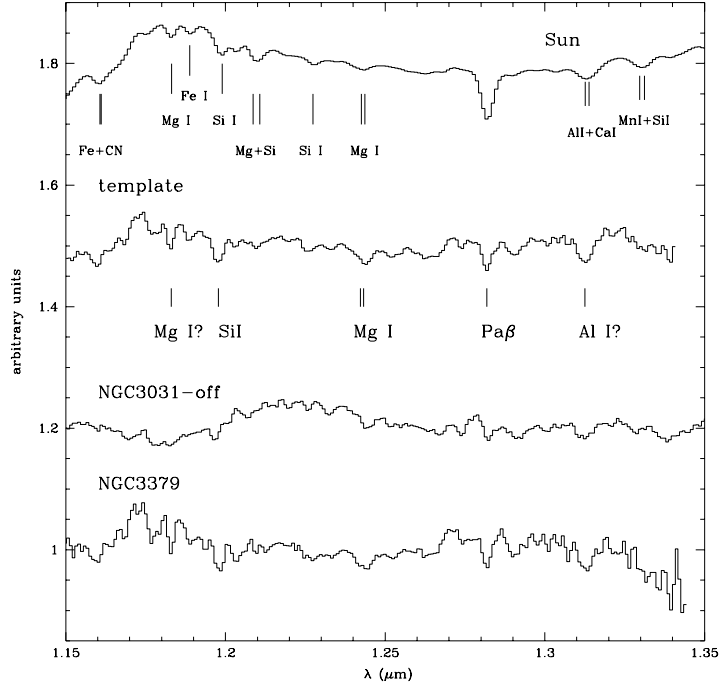}{425pt}{0}{50}{50}{-160}{70}
\vspace{-5cm}
\caption{From bottom to top, $J$-band
spectrum of the elliptical galaxy NGC~3379,
off-nucleus of NGC~3031, the result of combining the previous galaxies
(template spectrum), and the
sun (rebinned to our spectral resolution). The spectra
of NGC~3379, off-nucleus NGC~3031 and template have been shifted to the
rest-frame wavelength. We mark the absorption features identified in the
solar spectrum, and the tentative identifications in the
template spectrum.}
\end{figure*}

In Table~1 we present the sample of
galaxies: column~(3) gives the observing run for each galaxy, and
column~(4) the total integration time in seconds for the $J$-band spectra.
The integration times for the $H$ spectra are given in the
notes.

\begin{deluxetable}{lccccc}
\tablefontsize{\footnotesize}
\tablewidth{12cm}
\tablecaption{Near-infrared aperture photometry.}
\tablehead{\colhead{Galaxy} & \colhead{$t_{\rm int}$} &
\colhead{Diameter (\arcsec)} & \colhead{$J-H$} &
\colhead{$J-K$} & \colhead{$J$}}
\startdata
NGC~2639 & $J$ 540\,s & 3    & 0.96 & 1.23 & 12.85 \\
         & $H$ 480\,s & 6    & 0.93 & 1.22 & 11.85 \\
         & $K$ 600\,s & 9    & 0.92 & 1.18 & 11.34 \\
         &            & 12   & 0.93 & 1.19 & 11.01 \\
         &            & 15   & 0.93 & 1.19 & 10.77 \\
         &            & 30   & 0.93 & 1.19 & 10.15 \\
NGC~3367 & $J$ 480\,s & 3  & 0.80 & 1.22 & 14.02 \\
         & $H$ 600\,s & 6  & 0.85 & 1.14 & 13.14 \\
         & $K$ 600\,s & 9  & 0.84 & 1.10 & 12.66 \\
         &            & 12 & 0.83 & 1.08 & 12.34 \\
         &            & 15 & 0.83 & 1.06 & 12.12 \\
         &            & 30 & 0.77 & 1.01 & 11.30 \\
         &            & 45 & 0.74 & 1.00 & 10.79 \\
NGC~3504 & $J$ 600\,s & 3 & \nodata & \nodata & 12.16 \\
         &            & 6 & \nodata & \nodata & 11.17 \\
         &            & 9 & \nodata & \nodata & 10.81 \\
         &            & 12 & \nodata & \nodata & 10.61\\
         &            & 15 & \nodata & \nodata & 10.47 \\
         &            & 30 & \nodata & \nodata & 10.05 \\
NGC~3998 & $J$ 300\,s & 3 & \nodata  & 1.12 & 10.00 \\
         & $K$ 300\,s & 6 & \nodata  & 1.06 &  9.11 \\
         &            & 9 & \nodata  & 1.05 &  8.72 \\
         &            & 12 & \nodata & 1.04 &  8.48 \\
         &            & 15 & \nodata & 1.04 &  8.31 \\
         &            & 30 & \nodata & 1.02 &  7.93 \\
NGC~4579 & $J$ 300\,s  & 3 & \nodata & \nodata & 11.69 \\
         &             & 6 & \nodata & \nodata & 10.76 \\
         &             & 9 & \nodata & \nodata &  10.31 \\
         &             & 12 & \nodata & \nodata & 10.01 \\
         &             & 15 & \nodata & \nodata & 9.78 \\
         &             & 30 & \nodata & \nodata & 9.12 \\
NGC~5953 & $J$ 300\,s & 3 & \nodata  & 1.26 & 12.84 \\
         & $K$ 300\,s & 6 & \nodata  & 1.13 & 11.89 \\
         &            & 9 & \nodata  & 1.11 & 11.47 \\
         &            & 12 & \nodata & 1.09 & 11.21 \\
         &            & 15 & \nodata & 1.08 & 11.06 \\
         &            & 30 & \nodata & 1.05  & 10.73 \\
NGC~5954 & $J$ 300\,s & 3 & \nodata  & 1.29 & 14.75 \\
         & $K$ 300\,s & 6 & \nodata  & 1.14 & 13.70 \\
         &            & 9 & \nodata  & 1.11 & 13.13 \\
         &            & 12 & \nodata & 1.08 & 12.74 \\
         &            & 15 & \nodata & 1.06 & 12.48 \\
         &            & 30 & \nodata & 1.01 & 11.74 \\
NGC~7743 & $J$ 600\,s & 3  & 0.70 & 1.10 & 12.56 \\
         & $H$ 600\,s & 6  & 0.70 & 1.05 & 11.81 \\
         & $K$ 600\,s & 9  & 0.69 & 1.02 & 11.45 \\
         &            & 12 & 0.69 & 1.00 & 11.21 \\
         &            & 15 & 0.68 & 0.99 & 11.03 \\
         &            & 30 & 0.65 & 0.94 & 10.49 \\
\enddata
\end{deluxetable}

\subsection{Near-Infrared Imaging and Flux-Calibration of the IR Spectra}

Because of the small projected width of the
slit and the unknown slit losses, we flux calibrated the spectra with images. We obtained broad-band images through $J$, $H$ and $K{\rm s}$ filters with a NICMOS3 array at the Steward Observatory 90'' telescope on Kitt Peak for seven galaxies in our sample on the nights of 1997 January 15
(NGC~2639 and NGC~7743), 1997 May 24 (NGC~3367), 1998 January
14 (NGC~5953+NGC~5954 only in $J$ and $K$ and NGC~4957 only
in $J$), 1998 April 14 (NGC~3504 only $J$ band)
and 1998 November 6 (NGC~3998 $J$ and $K$). For all observing
runs the pixel size was 0\farcs6 pixel$^{-1}$. The data were reduced
following a standard procedure which basically involves dark subtraction, flat
fielding, shifting of the galaxy images to a common position and
median-combining to produce the final image. Standard stars from the
list of Elias et al. (1982) were observed several times
throughout the night to provide a photometric calibration for the 1997 January
and 1998 November data. We used the standard extinction coefficients for
the site;  errors from the photometric calibration are $\pm 0.06, 0.07$ and 0.06\,mag at $J$, $H$ and $K$ respectively. Conditions were non-photometric in the 1997 May, 1998 January and 1998 April runs,  so the images had to be calibrated with published fluxes. For NGC~3367 we used  the
15\arcsec-diameter photometry given in Spinoglio et al. (1995), for
the interacting pair NGC~5953+NGC~5954 we used
the 5.5\arcsec-diameter photometry from Bushouse \& Stanford (1992), for
NGC~4579 the 5.5\arcsec-diameter photometry of Devereux, Becklin \&
Scoville (1987), and finally for NGC~3504 the 10\arcsec-diameter
photometry from Balzano \& Weedman (1981).
The integration times for each filter along with the aperture
photometry for these galaxies are presented in Table~2.

\begin{figure*}
\figurenum{3}
\plotfiddle{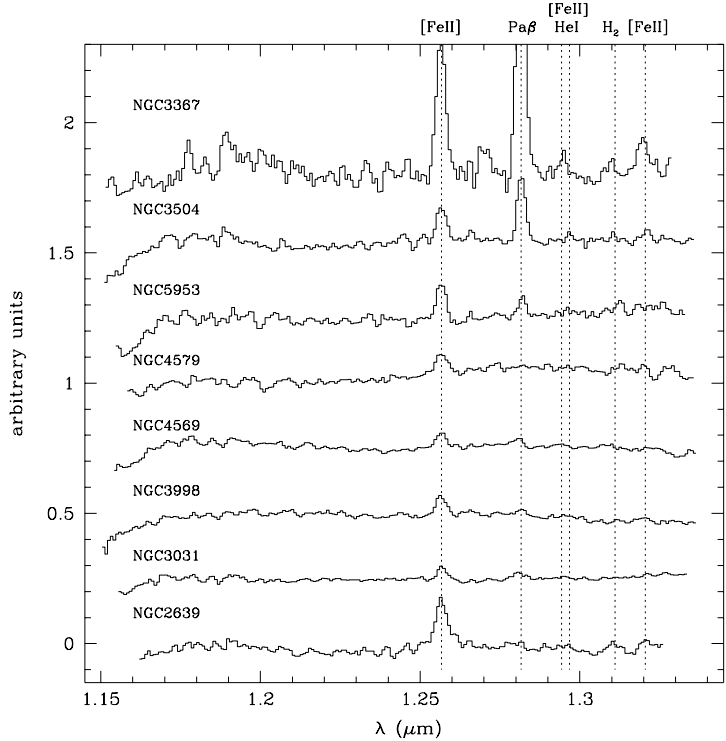}{425pt}{0}{50}{50}{-160}{90}
\vspace{-6.cm}
\caption{Stellar continuum subtracted spectra
($1.2\arcsec \times 2\arcsec$ aperture) in the $J$-band
shifted to the rest-frame wavelength. The spectra have been moved up
by an arbitrary amount for displaying purposes. We indicate the position of
[Fe\,{\sc ii}]$1.256\,\mu$m and Pa$\beta$. Also marked
possible detections of the  emission lines
[Fe\,{\sc ii}]$1.2943\,\mu$m and He\,{\sc
i}$\lambda 1.2968\,\mu$m, H$_2$(4,2)S(1)$\lambda 1.3112\,\mu$m and
[Fe\,{\sc ii}]$1.3205\,\mu$m.}
\end{figure*}

The flux-calibration of the near-IR spectra was performed using the
3\arcsec-diameter \  photometry from the images (Table~2), which
approximately corresponds to the region covered by the large extraction
aperture ($1.2\arcsec \times 6\arcsec$), except NGC~3504 for which we
used a $1.2\arcsec \times 8.5\arcsec$ aperture to match that of our
data in Alonso-Herrero et al. (1997), and
NGC~3998 for which a $1.2\arcsec \times 4\arcsec$ approximately
matches the aperture of Larkin et al. (1998). The small aperture spectra
($1.2\arcsec \times 2\arcsec$) were flux-calibrated relative to the
large aperture spectra.
The  spectra of NGC~3031, NGC~3504 and NGC~4569 were calibrated with
aperture photometry from the
literature (see references in Table~1, column~(7)).

\section{ANALYSIS}

In Figure~1 we present the fully reduced $J$-band spectra,
normalized to unity (at $1.25\,\mu$m). These spectra were extracted
with an aperture of $1.2 \arcsec \times 2\arcsec$. The
[Fe\,{\sc ii}]$1.2567\,\mu$m line is clearly detected in all cases, whereas
the Pa$\beta$ line is not detected in NGC~3998,
NGC~7743, and is probably seen in absorption in NGC~2639, NGC~3031,
NGC~4569 and NGC~4579. For the galaxies in which Pa$\beta$ is not detected,
its expected location is indicated, inferred from the observed
wavelength of [Fe\,{\sc ii}]$1.2567\,\mu$m. Table~3 lists for each galaxy the line measurements for the apertures used to flux calibrate the spectra (given in the second column) in the first row, whereas the second row gives the
same quantities for a $1.2 \arcsec \times 2\arcsec$ aperture.
When Pa$\beta$ was not observed in emission an upper limit
is tabulated. The full widths at half maximum (FWHM) are corrected for the
instrument resolution (${\rm FWHM }_{\rm inst} \simeq 650\,$km s$^{-1}$ at
$1.25\,\mu$m, as derived from the sky line widths). The $J$-band emission
lines of NGC~3031, NGC~3367, NGC~3504, NGC~4569 and NGC~7743 appear unresolved
at our resolution, while the other galaxies have marginally resolved lines.
Due to the uncertainties of the continuum placement, the line fluxes,
particularly of Pa$\beta$, are uncertain. For the two galaxies in common with
Larkin et al. (1998), we measure a [Fe\,{\sc ii}]$1.257\,\mu$m flux
three times as large as they do for NGC 3998, whereas
given the differing aperture sizes, the
[Fe\,{\sc ii}]$1.257\,\mu$m fluxes for NGC 7743 are consistent.

\begin{deluxetable}{lccccccc}
\tablefontsize{\footnotesize}
\tablecaption{$J$-band spectral features.}
\tablehead{\colhead{Galaxy} & \colhead{Aperture} &
\multicolumn{3}{c}{[Fe\,{\sc ii}]$1.257\,\mu$m} &
\multicolumn{3}{c}{Pa$\beta$} \\
& & \colhead{flux} &\colhead{EW} & \colhead{FWHM} & \colhead{flux} &
\colhead{EW} & \colhead{FWHM} \\
\colhead{(1)} & \colhead{(2)} & \colhead{(3)} & \colhead{(4)} &
\colhead{(5)} & \colhead{(6)} & \colhead{(7)} & \colhead{(8)}}
\startdata
NGC~2639 & $1.2\arcsec \times 6\arcsec$ & $1.38\times 10^{-14}$ &
6.1 & 675 & $<2\times 10^{-15}$ & \nodata & \nodata \\
& $1.2\arcsec \times 2\arcsec$ &  $1.14\times 10^{-14}$ &
8.0 & 675 & $<1\times 10^{-15}$ & \nodata & \nodata \\
NGC~3031 & $1.2\arcsec \times 6\arcsec$ & $2.73\times 10^{-14}$ &
1.1 & $<650$ & $<1\times 10^{-14}$ & \nodata & \nodata \\
& $1.2\arcsec \times 2\arcsec$ &  $1.80\times 10^{-14}$ &
1.5 & $<650$ & $<6\times 10^{-15}$ & \nodata & \nodata \\
NGC~3367 & $1.2\arcsec \times 6\arcsec$ & $1.28\times 10^{-14}$ &
16.0 & $<650$ & $2.11\times 10^{-14}$ & 26.9 & $<650$ \\
& $1.2\arcsec \times 2\arcsec$ &  $8.18\times 10^{-15}$ &
21.6 & $<650$ & $1.34\times 10^{-14}$ & 33.5 & $<650$ \\
NGC~3504 & $1.2\arcsec \times 8.5\arcsec$ & $3.80\times 10^{-14}$ &
5.3 & $<650$ & $8.03 \times 10^{-14}$ & 11.1 & $<650$\\
& $1.2\arcsec \times 2\arcsec$ &  $1.45\times 10^{-14}$ &
4.3 & $<650$ & $2.42\times 10^{-14}$ & 7.2 & $<650$ \\
NGC~3998 & $1.2\arcsec \times 4\arcsec$ &  $1.60 \times 10^{-14}$ &
2.5 & $<650$ & $<4\times 10^{-15}$ & \nodata & \nodata \\
& $1.2\arcsec \times 2\arcsec$ & $1.10 \times 10^{-14}$ &
2.9 & $<650$ & $<4\times 10^{-15}$ & \nodata & \nodata \\
NGC~4569 & $1.2\arcsec \times 8.5\arcsec$ & $1.70\times 10^{-14}$ &
1.4 & $<650$ & $<7 \times 10^{-15}$ & \nodata &\nodata \\
& $1.2\arcsec \times 2\arcsec$ &  $1.42\times 10^{-14}$ &
1.8 & $<650$ & $<3\times 10^{-15}$ & \nodata & \nodata \\
NGC~4579  & $1.2\arcsec \times 6\arcsec$ & $1.76\times 10^{-14}$&
2.6 & 875 & $<5\times 10^{-15}$ & \nodata & \nodata \\
& $1.2\arcsec \times 2\arcsec$ & $1.20\times 10^{-14}$&
3.4 & 875 & $<3\times 10^{-15}$  & \nodata & \nodata \\
NGC~5953 & $1.2\arcsec \times 6\arcsec$ & $1.10\times 10^{-14}$ &
4.8 & $<650$ & $7.97\times 10^{-15}$  & 2.8  & $<650$ \\
& $1.2\arcsec \times 2\arcsec$ & $6.70\times 10^{-15}$ &
4.7 & $<650$ & $3.05\times 10^{-15}$  & 2.3  & $<650$ \\
NGC~7743 & $1.2\arcsec \times 6\arcsec$ & $5.63\times 10^{-15}$ &
3.2 & $<650$ & $<9 \times 10^{-16}$ & \nodata & \nodata \\
\enddata
\tablecomments{Col.~(1) Galaxy. Col.~(2) Extraction aperture. Col.~(3) and
(6) Fluxes
in erg cm$^{-2}$ s$^{-1}$ measured from observed
spectra (before stellar continuum subtraction).
Col.~(4) and (7) EW in \AA. Col.~(5) and
(8) FWHM in km s$^{-1}$ (corrected for instrumental resolution).}
\end{deluxetable}

\begin{deluxetable}{lcccc}
\tablefontsize{\footnotesize}
\tablecaption{Absorption features observed in the $J$-band
template spectrum.}
\tablehead{\colhead{$\lambda_{\rm obs}$} & \colhead{Species} &
\colhead{$\lambda_{\rm vac}$} & \colhead{Galaxies} &\colhead{Comment}\\
\colhead{(1)} &\colhead{(2)} &\colhead{(3)} &\colhead{(4)} &\colhead{(5)}}
\startdata
$1.1828 \pm 0.004\,\mu$m & Mg\,{\sc i} &
$1.1828\,\mu$m & NGC~2639, NGC~3031, NGC~3504 ? & a\\
& & & NGC~4569, NGC~5953\\
$1.1885 \pm 0.004\,\mu$m & Fe\,{\sc i} &
$1.1886\,\mu$m & NGC~2639 ?, NGC~3031, NGC~4569 & a\\
$ 1.1978 \pm 0.004\,\mu$m &  Si\,{\sc i}  &
$1.1984, 1.1991\,\mu$m & NGC~2639, NGC~3031, NGC~3504? \\
            & Fe\,{\sc i} & $1.1976\,\mu$m & NGC~3998, NGC~4569, NGC~4579
& a \\
$1.2433 \pm 0.004\,\mu$m & Mg\,{\sc i}  &
$1.2424, 1.2433\,\mu$m & NGC~2639, NGC~3031, NGC~3998 & b \\
&K\,{\sc i} & $1.2432\,\mu$m & NGC~4569, NGC~4579, NGC~5953 & c, d\\
$1.3126 \pm 0.004\,\mu$m & Al\,{\sc i} &
$1.3123\,\mu$m & NGC~3031, NGC~3504?, NGC~3367 ? & \\
& Ca\,{\sc i} & $1.3135\,\mu$m & NGC~3998?, NGC~4569  & d\\
\enddata
\tablecomments{Col.~(1) Observed wavelength of the feature in the
template spectrum. Col.~(2) and (3) Tentative identification and
vacuum wavelength. Col.~(4) Galaxies in which the feature is tentatively
detected. Col.~(5) Comments.\\
a. The Mg\,{\sc i}$\lambda 1.1828\,\mu$m, and the
Fe\,{\sc i} lines  have been identified in the spectra of cool low-mass stars
(Jones et al. 1996).\\
b. The Mg\,{\sc i} doublet is identified in LINERs (Larkin et al. 1998).\\
c. The K\,{\sc i} line is present in M dwarf stars (Kirkpatrick et al. 1993).\\
d. The Ca\,{\sc i} line is identified in M stars (Jones et al. 1994).\\}
\end{deluxetable}

\subsection{Stellar continuum subtraction}

The strong stellar continuum dominates the $J$-band spectra of all our
LINERs but NGC~3367. To improve the estimates of line strengths, we have used
a "template" spectrum to subtract this continuum. Our template is produced
from a spectrum of the elliptical galaxy NGC~3379 combined with our
off-nuclear spectrum of M81. The former galaxy is often used
for similar purposes in the optical, and the latter spectrum samples the
stellar population actually present close to a LINER nucleus.
The spectra are shown at the bottom of Figure~2. These spectra
have been normalized by fitting the slope of the continuum.
Within the errors, there is no difference in the two template components. The Pa$\beta$ line is observed in absorption with equivalent width ${\rm EW} =
-1.6\,$\AA \ together with a number of absorption features (see below).
We also display the solar spectrum\footnote{NSO/Kitt Peak FTS data used here
were produced by NSF/NOAO} between 1.15 and $1.35\,\mu$m, rebinned to
our spectral resolution (top spectrum in Figure~2).  We give possible
identifications from Livingston \& Wallace (1991) and Striganov \&
Sventinskii (1968) for some of the absorptions, as indicated in
Figure~2 and summarized in Table~4. The wavelengths quoted are the average
of the measured wavelengths in all the spectra where the feature is observed
(the galaxies in which a given feature is seen are listed in the last column).

\begin{figure*}
\figurenum{4}
\plotfiddle{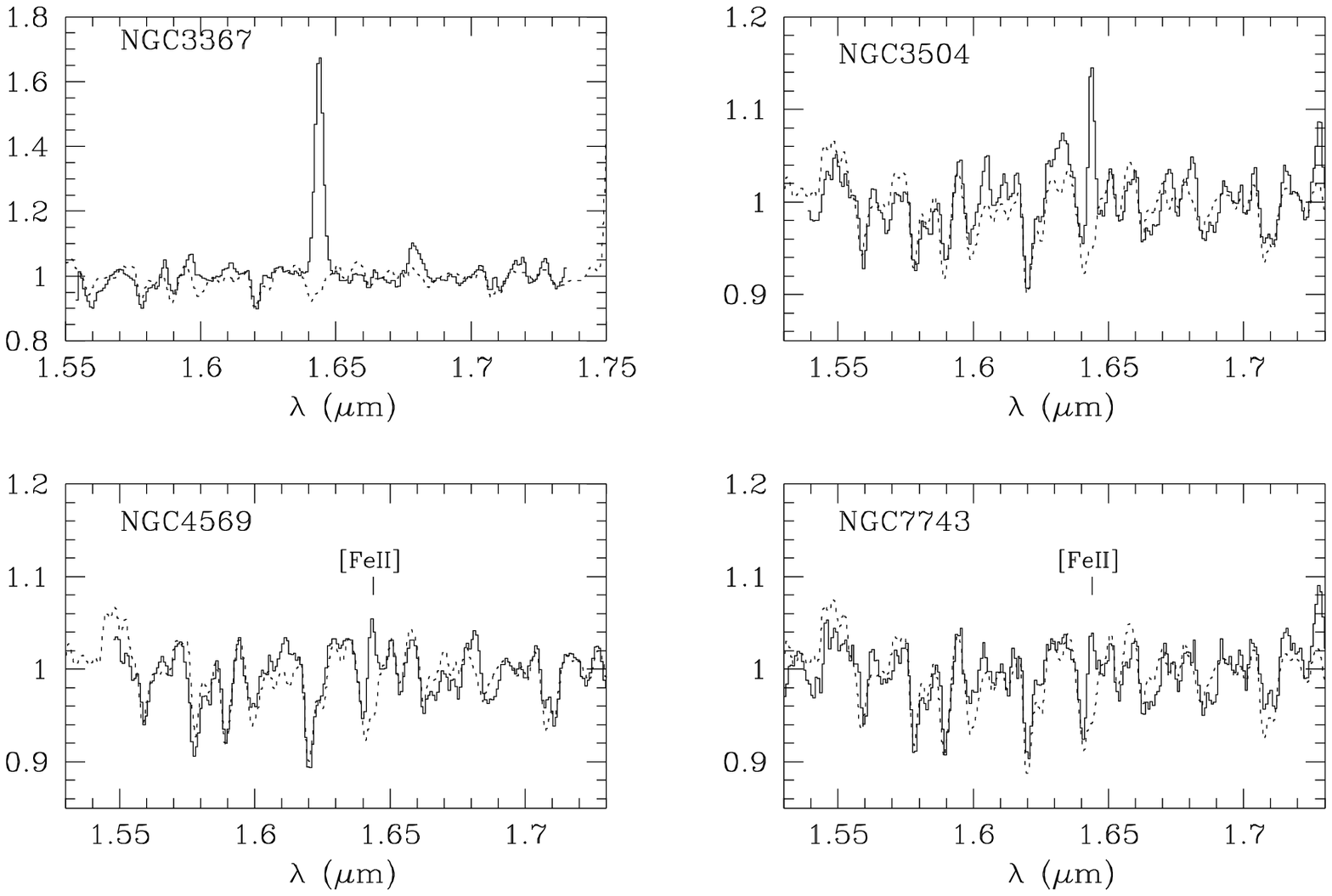}{425pt}{0}{60}{60}{-160}{-60}
\vspace{-4cm}
\caption{Observed $H$-band spectra of four LINERs (solid
line) normalized to unity
extracted with a $1.2\arcsec \times 2\arcsec$ aperture. For each
galaxy we plot as dotted lines the stellar template (composed of
M-type supergiants) diluted by a small factor (see text for
details).}
\end{figure*}

We shifted the LINER spectra to the
rest wavelength, normalized them to unity, and then subtracted the $J$-band
template. The stellar continuum subtracted spectra of our LINERs
are presented in Figure~3, corresponding to spectra extracted with a $1.2\arcsec \times 2\arcsec$ aperture. Due to the low signal-to-noise  of the spectrum of
NGC~7743, we did not perform stellar continuum subtraction for it. We used its own off-nucleus spectrum only as the template for NGC~3031. The resulting
continuum-subtracted spectra are very flat with most of the absorption
features successfully removed, suggesting that the template matches quite well
the underlying stellar emission of LINERs. The removal of the underlying
stellar continuum allows the detection of weak emission lines. From
Figure~3, it can be seen that now Pa$\beta$ appears in emission in all the galaxies, except NGC~2639 and NGC~4579, where the detection is only
marginal. For NGC~3367 and NGC~3504, we also detect additional lines at rest wavelengths (marked in Figure~3): $\lambda = 1.296 \pm 0.001\,\mu$m (possible identifications [Fe\,{\sc ii}]$1.2943\,\mu$m and He\,{\sc
i}$\lambda 1.2968\,\mu$m), $\lambda = 1.311 \pm 0.001\,\mu$m
(H$_2$(4,2)S(1)$\lambda 1.3112\,\mu$m) and $\lambda = 1.321 \pm 0.001\,\mu$m
([Fe\,{\sc ii}]$1.3205\,\mu$m). These lines are also reported in the starburst
galaxy M82 (McLeod et al. 1993).

In Table~5 we give the [Fe\,{\sc ii}]$1.257\,\mu$m/Pa$\beta$ line ratios
for large apertures (typically the apertures used to flux-calibrate
the data, except for NGC~3998 where we used
a $1.2\arcsec \times 4\arcsec$ aperture), small apertures ($1.2\arcsec
\times 2\arcsec$), and that measured from the continuum subtracted spectra.
We also compare the continuum-subtracted Pa$\beta$
fluxes with the Pa$\beta$ fluxes predicted from H$\alpha$ fluxes
(not corrected for absorption) measured through apertures of similar area to
that given in column~(7) of Table~5. In general the agreement between
the measured Pa$\beta$ fluxes and those
predicted from the optical data is reasonably good.

\begin{figure*}
\figurenum{5}
\plotfiddle{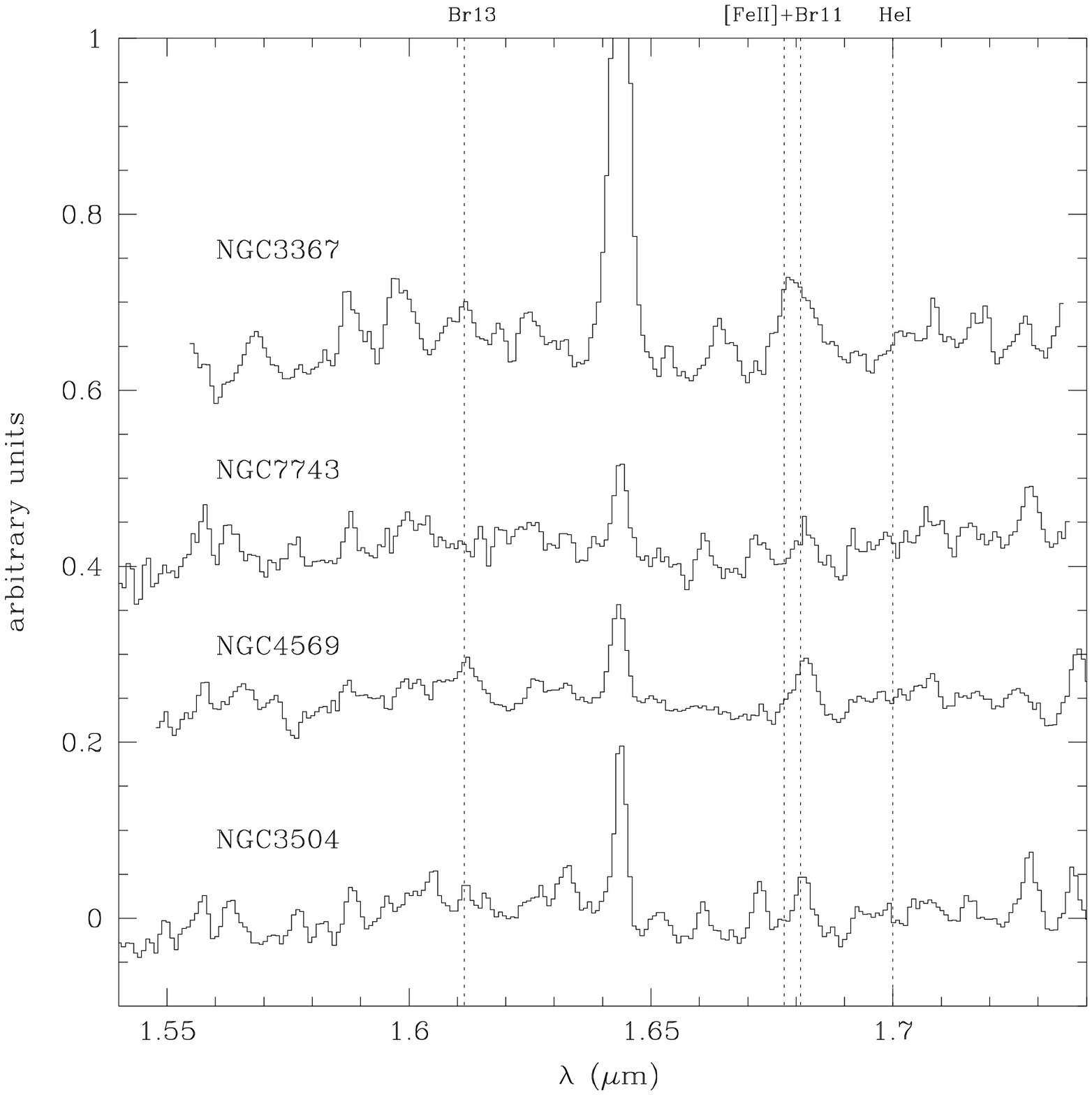}{425pt}{0}{40}{40}{-270}{150}
\plotfiddle{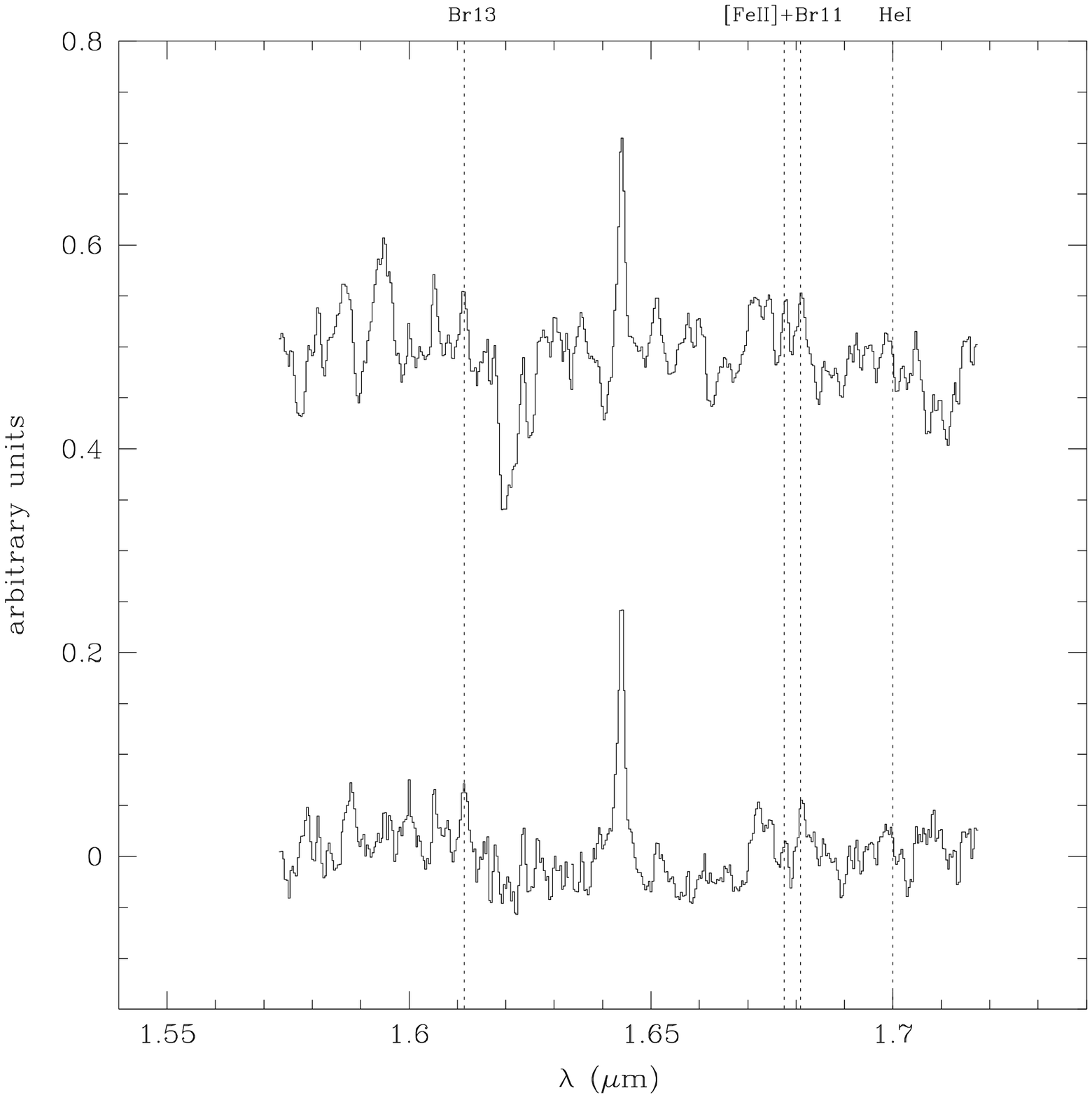}{425pt}{0}{40}{40}{0}{590}
\vspace{-22cm}
\caption{{\it Left panel}. Stellar continuum subtracted spectra in the $H$-band
shifted to rest frame wavelength. The spectra have been shifted vertically
by arbitrary amounts for displaying purposes.
We mark the expected positions of the emission lines Br13,
[Fe\,{\sc ii}]$1.677\,\mu$m, Br11 and He\,{\sc i}$1.70\,\mu$m.
Br11 is observed blended with [Fe\,{\sc ii}]$1.677\,\mu$m in the intermediate
resolution spectra, whereas both lines are clearly separated in the
high resolution spectrum of NGC~3504. The He\,{\sc i} is not clearly
detected. {\it Right panel}.Observed high resolution spectra (top) 
at $1.63\,\mu$m and
$1.68\,\mu$m of NGC~3504, and stellar continuum subtrated spectra (bottom).}
\end{figure*}

A similar procedure was applied to the $H$-band spectra (calibrated spectra
are presented in Figure~4, solid lines), but in this case we synthesized
the normal galaxy spectrum using an $H$-band stellar library
(Engelbracht, private communication). As for the $J$-band data, before the stellar continuum subtraction, the galaxy spectra were shifted to rest frame wavelength, and the continuum normalized to unity. We subtracted different empirical stellar templates (composed of different spectral types and luminosities) and found that the best result for all galaxies
was obtained with a combination of M supergiants. However, the stellar features had to be diluted to fit the galaxy absorption features. The dilution factors used for each galaxy are: 0.5 for NGC~3367 and NGC~4569, and 0.7 for NGC~3504 and NGC~4569. The stellar templates are plotted for each galaxy as dotted lines in Figure~4. It is likely that the spectra are dominated by AGB stars and perhaps by relatively low mass supergiants; the dilution of the composite supergiant spectrum is an {\it ad hoc} way to achieve a good fit, until a true population synthesis can be performed. The stellar continuum subtracted spectra of NGC~3367, NGC~3504, NGC~4569 and NGC~7743 in $H$ are presented in Figure~5a (intermediate resolution), whereas in Figure~5b we plot the high resolution spectra of NGC~3504 together with the stellar continuum subtracted data. Again the spectra in Figures~5a and 5b correspond to the spectra
extracted with a $1.2\arcsec \times 2\arcsec$; from these two figures
it can be seen that  most of the absorption features
have been successfully removed. Note in Figure~4 the presence of a strong
CO absorption feature near the [Fe\,{\sc ii}]$1.644\,\mu$m emission line,
which effectively dilutes the latter line in NGC~4569 and NGC~7743.
In Figure~5a we see how the [Fe\,{\sc ii}] line, barely seen in NGC~4569 and NGC~7743 (Figure~4), is recovered in the stellar continuum subtracted spectra.

\begin{deluxetable}{lcccccc}
\tablefontsize{\footnotesize}
\tablecaption{[Fe\,{\sc ii}]$1.257\,\mu$m/Pa$\beta$ line ratios
and Pa$\beta$ fluxes.}
\tablehead{\colhead{} & \multicolumn{3}{c}{[Fe\,{\sc ii}]/Pa$\beta$} &
\multicolumn{2}{c}{$f({\rm Pa}\beta)$}
& \colhead{Optical}\\
\colhead{Galaxy}& \colhead{large} & \colhead{small} & \colhead{subtracted} &
\colhead{measured} & \colhead{predicted} & \colhead{Aperture}\\
\colhead{(1)} &\colhead{(2)} &\colhead{(3)} &\colhead{(4)} &\colhead{(5)}
&\colhead{(6)} & \colhead{(7)}}
\startdata
NGC~2639 & $>6.9$ & $>7.8$ & $13.0\pm1$ & $1.10 \times 10^{-15}$ &
$2.36 \times 10^{-15}$ & $1\arcsec \times 4\arcsec$ \\
NGC~3031 & $>2.7$ & $>3.0$ & $1.50\pm 0.50$ & $1.80 \times 10^{-14}$
& $1.14 \times 10^{-14}$ & $1\arcsec \times 4\arcsec$\\
NGC~3367 & 0.61 & 0.59 & $0.58\pm0.05$ & $2.21 \times 10^{-14}$
& $1.16 \times 10^{-14}$ & ? \\
NGC~3504 & 0.47 & 0.60 & $0.54\pm0.05$ & $5.20 \times 10^{-14}$
& $1.89 \times 10^{-14}$ & $1\arcsec \times 4\arcsec$ \\
NGC~3998 & $>4.0$  & $>3.5$ & $3.45\pm 0.30$ & $4.10 \times 10^{-15}$
& $6.25 \times 10^{-15}$ & $1\arcsec \times 4\arcsec$ \\
NGC~4569 & 2.4 & 3.5 & $1.50\pm0.40$ & $1.13 \times 10^{-14}$
& $2.08 \times 10^{-14}$ & round 2\arcsec \\
NGC~4579 & $>3.2$   & $>4.0$ & $3.30\pm0.30$ &$3.60\times 10^{-15}$
& $2.78 \times 10^{-15}$
& round 2\arcsec \\
NGC~5953 & 1.40  & 2.20 & $1.90\pm0.20$ &$5.80 \times 10^{-15}$
& $1.02 \times 10^{-14}$ & ? \\
NGC~7743 & \nodata & $>6.3$ & \nodata & $<9 \times 10^{-16}$
& $2.64 \times 10^{-15}$  & $2\arcsec \times 4\arcsec$\\
\enddata
\tablecomments{Col.~(1) Galaxy. Col.~(2)--(4)
[Fe\,{\sc ii}]$1.257\,\mu$m/Pa$\beta$ line ratio measured from the
spectra extracted with the  large and small aperture, and from the
stellar continuum subtracted spectrum (small aperture).
Col.~(5) Pa$\beta$ line fluxes
(in erg cm$^{-2}$ s$^{-1}$) measured from
the stellar continuum subtracted spectra with apertures
similar to optical spectroscopy, column~(7). When the optical
aperture is not given, the Pa$\beta$ flux corresponds to the
$1.2\arcsec \times 2\arcsec$ aperture. Col.~(6) Pa$\beta$ line fluxes
(in erg cm$^{-2}$ s$^{-1}$) predicted from the H$\alpha$ fluxes.
Col.~(7) aperture for data in Col.~(5) and (6).}
\end{deluxetable}

\begin{deluxetable}{lcccccccc}
\tablefontsize{\footnotesize}
\tablecaption{$H$-band spectral features.}
\tablehead{\colhead{} & \colhead{} &
\multicolumn{3}{c}{[Fe\,{\sc ii}]1.644+Br12} &
 & & Ratio [Fe\,{\sc ii}] \\
\colhead{Galaxy} & \colhead{Aperture} &
\colhead{flux} &
\colhead{EW} & \colhead{FWHM} &
\colhead{$f({\rm Br}12)_{\rm pred}$} &
\colhead{$f({\rm Br}11)_{\rm pred}$} &
\colhead{1.257/1.644}\\
\colhead{(1)} &\colhead{(2)} &\colhead{(3)} &\colhead{(4)} &
\colhead{(5)} &\colhead{(6)} &\colhead{(7)} &\colhead{(8)}}
\startdata
NGC~3367 &$1.2\arcsec \times 6\arcsec$ &
$1.19\times 10^{-14}$ & 19.2 & $<400$ & $6.8\times 10^{-16}$
& $8.8 \times 10^{-16}$ & 1.1\\
& $1.2\arcsec \times 2\arcsec$ & $9.34\times 10^{-15}$ & 23.0 & $<400$
& \nodata & \nodata & \nodata \\

NGC~3504 &$1.2\arcsec \times 8.5\arcsec$ &
$2.45\times 10^{-14}$ &  4.7 & $<400$ & $2.6\times 10^{-15}$ &
$3.4\times 10^{-15}$ & 1.7 \\
& $1.2\arcsec \times 2\arcsec$ & $1.05\times 10^{-14}$ & 4.1 & $<400$
& \nodata & \nodata & \nodata \\
& $1.2\arcsec \times 2\arcsec$\tablenotemark{a} &  $2.30\times 10^{-14}$ &
3.8 & 200 &\nodata & $5.1\times 10^{-15}$\tablenotemark{b} \\
NGC~4569 &$1.2\arcsec \times 8.5\arcsec$
& $1.40\times 10^{-14}$ & 1.5 &  $<400$ & $2.2\times 10^{-16}$ &
$2.9\times 10^{-16}$ & 1.2 \\
& $1.2\arcsec \times 2\arcsec$ & $6.84\times 10^{-15}$ & 1.2 & $<400$
& \nodata & \nodata & \nodata \\
NGC~7743 &$1.2\arcsec \times 6\arcsec$
& $<4.3\times 10^{-15}$ & \nodata & \nodata &
$2.9\times 10^{-17}$ & $3.8\times 10^{-17}$ & 1.3 \\
\enddata
\tablecomments{Col.~(1) Galaxy. Col.~(2) Extraction aperture. Col.~(3) Flux
in erg cm$^{-2}$ s$^{-1}$. Col.~(4) EW in \AA. Col.~(5) FWHM in
km s$^{-1}$ (corrected for instrumental resolution). Col.~(6) Flux of Br12
(erg cm$^{-2}$ s$^{-1}$) computed from Pa$\beta$. Col.~(7) Flux of Br11
(erg cm$^{-2}$ s$^{-1}$) computed from Pa$\beta$.  Col.~(8)
[Fe\,{\sc ii}]$1.257\,\mu$m/[Fe\,{\sc ii}]$1.644\,\mu$m line ratio.}
\tablenotetext{a}{High-resolution data}
\tablenotetext{b}{Measured}
\end{deluxetable}

The He\,{\sc i}($1.70\,\mu$m)Br10 line ratio is
very sensitive to the stellar temperature and almost reddening and
electron temperature independent, and it is found to be
He\,{\sc i}/Br$10 \simeq 0.35$ for stellar temperatures of
40,000\,K (Vanzi et al. 1996). The main purpose of the continuum subtraction in the $H$-band is to detect (or set an upper limit) to the fluxes
of Br11 (and Br10) and He\,{\sc i} $1.70\,\mu$m. To help with the
identifications, we mark the vacuum wavelengths of Br13, Br11, 
[Fe\,{\sc ii}]$1.677\,\mu$m and He\,{\sc i} $1.70\,\mu$m in 
Figures~5a and 5b. In the intermediate resolution data, the Br11 
line appears to be blended
with the [Fe\,{\sc ii}]$1.677\,\mu$m line, whereas the He\,{\sc i} 
is not detected. The high resolution spectra of NGC~3504 show the 
Br11 and [Fe\,{\sc ii}]$1.677\,\mu$m resolved (Figure~5b). In Table~6
we give the flux, equivalent width and FWHM (corrected for the
instrument resolution) of the [Fe\,{\sc
ii}]$1.644\,\mu$m+Br12 lines, together with the predicted fluxes for the Br11
and Br12 lines from the observed values of Pa$\beta$ (and from
Br$\gamma$ when available). The last column gives
the line ratio [Fe\,{\sc ii}]$1.257\,\mu$m/[Fe\,{\sc ii}]$1.644\,\mu$m for the
four galaxies as measured from stellar continuum subtracted
spectra; the theoretical value for this line ratio is 1.36 and is
independent of the physical conditions (Nussbaumer \& Storey 1988).
The Br11  (at $1.681\,\mu$m) line fluxes
for NGC~3504 and NGC~3367 are measured from the stellar-continuum
subtracted spectra, and converted into Br10 fluxes using the
theoretical line ratio ${\rm Br10}/{\rm Br11}=1.33$ (Hummer \&
Storey 1987). Our upper limit to the
line flux of the He\,{\sc i} line leads to He\,{\sc i}/Br$10 < 0.26$
in both NGC~3504 and NGC~3367 and  places an upper limit of $\simeq 40,000\,$K
on the  stellar temperature.

\subsection{LINER Classification from the [Fe\,{\sc
ii}]$1.257\,\mu$m/Pa$\beta$ line ratio}

The [Fe\,{\sc ii}]$1.257\,\mu$m/Pa$\beta$ (or [Fe\,{\sc
ii}]$1.644\,\mu$m/Br$\gamma$) line ratio used in conjunction with [O\,{\sc
i}]$\lambda6300$/H$\alpha$ has proven to be useful in separating different
types of activity (Mouri et al. 1990; Goodrich, Veilleux, \&
Hill 1994; Simpson et al. 1996; Alonso-Herrero et al. 1997; Veilleux,  Goodrich, \& Hill, 1997). In Alonso-Herrero et al. (1997) we showed that the behavior of both the near-infrared and the optical line ratios can be understood as a progression in the proportion of shock excitation going from H\,{\sc ii} regions, through starbursts and Seyferts to SNRs. In Figure~6 we show an
[Fe\,{\sc ii}]$1.257\,\mu$m/Pa$\beta$ vs. [O\,{\sc  i}]$\lambda6300$/H$\alpha$
diagram where we plot data for our sample of LINERs, together with the
line ratios of NGC~1052 and NGC~253. In addition we show Larkin et al.'s
(1998) sample of LINERs as filled squares, which represent the near-IR line
ratio obtained by them from the Pa$\beta$ flux predicted from H$\alpha$.
The boxes indicate the approximate locations of starburst and
Seyfert galaxies.

\begin{figure*}
\figurenum{6}
\plotfiddle{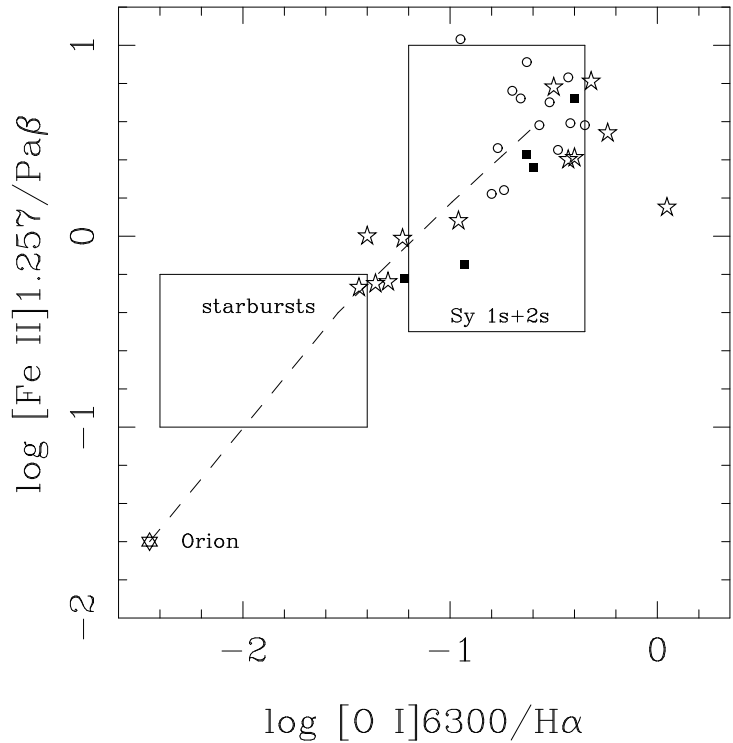}{425pt}{0}{80}{80}{-160}{130}
\vspace{-5cm}
\caption{[Fe\,{\sc ii}]$1.257\,\mu$m/Pa$\beta$ vs.
[O\,{\sc  i}]$\lambda6300$/H$\alpha$ diagram. The star symbols are our
sample of galaxies (including NGC~1052 and NGC~253), the filled squares
are Larkin et al. (1998) LINERs. The open circles are
SNRs from Lumsden \& Puxley (1995). The dashed line represents
a mixing curve of a pure H\,{\sc ii} region (e.g., Orion) and
SNRs. The boxes indicate the approximate location of the regions occupied
by starbursts and Seyfert galaxies.}
\end{figure*}

From the same type of diagram Larkin et al. (1998) concluded that there are
two classes of LINERs, weak-[Fe\,{\sc ii}] LINERs ([Fe\,{\sc
ii}]$1.257\,\mu$m/Pa$\beta < 2$) and strong-[Fe\,{\sc ii}] LINERs
([Fe\,{\sc ii}]$1.257\,\mu$m/Pa$\beta > 2$); according to these
authors the former class would be low luminosity AGNs, whereas the latter
class of LINERs would be powered by starbursts. However, we could easily reach the opposite conclusion. Four objects in our sample appear to be (or have been)
dominated by star formation, NGC~3504 (Alonso-Herrero et al. 1997 and
references therein), NGC~253 (Engelbracht et al. 1998), NGC~3367 (there is
evidence for the presence of Wolf-Rayet stars, Ho et al. 1995), and
NGC~4569 (Keel 1996; Maoz et al. 1998); all these objects show [Fe\,{\sc
ii}]$1.257\,\mu$m/Pa$\beta <0.70$, and can be classified as weak-[O\,{\sc i}]
LINERs. Another group of LINERs shows evidence for having an AGN, mainly based
on the detection of a broad component in H$\alpha$:
NGC~3031 (Filippenko \& Sargent 1988), NGC~2639 and NGC~3998
(Filippenko \& Sargent 1985), and NGC~4579 (Ho et al. 1995). All these
galaxies have [Fe\,{\sc ii}]$1.257\,\mu$m/Pa$\beta > 1.40$.
If now we turn to Larkin et al. (1998) sample,
we find that NGC~404 and NGC~4736 may have had
recent star-formation (Maoz et al. 1998;
Ho et al. 1993, 1995, and Walker et al. 1988 and Taniguchi et al. 1996
respectively), yet [Fe\,{\sc ii}]$1.257\,\mu$m/Pa$\beta > 3$. NGC~4258
has broad H$\alpha$ (Filippenko \& Sargent 1985), and NGC~5194 is
classified as intermediate between LINER and Seyfert; both have
[Fe\,{\sc ii}]$1.257\,\mu$m/Pa$\beta = 0.6 - 2$. NGC~4826
has optical line ratios (Keel 1983) which locate it in the LINER region in
diagnostic diagrams (Veilleux \& Osterbrock 1987; Ho et al.
1997a), but it is a weak-[O\,{\sc i}] LINER
with [Fe\,{\sc ii}]$1.257\,\mu$m/Pa$\beta = 0.7$. Distinguishing different
types of LINERs solely from the [Fe\,{\sc ii}]$1.257\,\mu$m/Pa$\beta$ ratio
appears to be problematic.

\subsection{LINERs powered by an evolved starburst}

Although AGN excitation can be identified in some LINERs through
the detection of a broad component of H$\alpha$, we believe many other LINERs are powered by an aging starburst. We find candidates for the latter category among weak-[O\,{\sc i}] LINERs and starburst/LINER transition objects as well as some {\it pure} LINERs (i.e., those which satisfy the Heckman (1980)
LINER definition). 

The emission-line behavior of starbursts can be
understood as a transition from pure H\,{\sc ii} region in the first
$\sim$ 10 million years, to an increasing role of shock excitation by supernova remnants, possibly ending with a pure supernova-like shock excited spectrum (Rieke et al. 1993; Genzel et al. 1995; Engelbracht et al. 1996, 1998; B\"oker, F\"orster-Schreiber, \& Genzel 1997; Vanzi et al. 1998). This transition occurs because the hot, ionizing stars in a starburst stellar
population have lifetimes of only a few million years, while supernovae are
produced by stars that are of relatively low mass and are only relatively weak
UV emitters (see, e.g., the starburst models of Leitherer \& Heckman 1995). 

We propose that traditional weak-[O\,{\sc i}] LINERs and
H\,{\sc ii}/LINERs arise as the episode of star formation ages and the SNRs start making an important contribution to the observed flux of the [Fe\,{\sc ii}] lines, as well as to the low-ionization optical emission lines. The [Fe\,{\sc ii}]$1.257\,\mu$m/Pa$\beta$ line ratio will rise as the number of SNR increases and the ionization flux from very young stars decreases (and therefore the Pa$\beta$ flux). This hypothesis has been proven to be successful in explaining the transitional LINER/H\,{\sc ii} line ratios of NGC~253 (Engelbracht et al. 1998).

Simpson et al. (1996) advanced  the idea that the
[Fe\,{\sc ii}]$1.257\,\mu$m/Pa$\beta$ line ratio could be used as an
indicator of the age of the starburst. Colina (1993) set an upper
limit to the [Fe\,{\sc ii}]$1.644\,\mu$m/Br$\gamma$ line ratio
(1.42 or [Fe\,{\sc ii}]$1.257\,\mu$m/Pa$\beta = 0.33$) produced
by star formation processes. However, Vanzi \& Rieke (1997)
calibrated the [Fe\,{\sc ii}] emission in terms of the SN rate by means
of M82, and showed that when this relationship is combined with an
evolutionary synthesis model one can reproduce a variety of
[Fe\,{\sc ii}]$1.644\,\mu$m/Br$\gamma$ line ratios from
$\simeq 2 \times 10^{-3}$ for very young starbursts up to
10 for evolved starbursts. The latter ratio is considerably higher
than the upper limit imposed by Colina's (1993) models.
In Figure~7 we present a plot of [Fe\,{\sc ii}]$1.644\,\mu$m/Br$\gamma$
versus $\log(K/{\rm Br}\gamma)$. The latter quantity is proportional to the
inverse of the equivalent width of Br$\gamma$ and is expressed in units of
$\mu{\rm m}^{-1}$. We  show in this diagram the LINERs believed
to be dominated by star formation  from our sample and
other references as filled squares and star-like symbols
respectively. The AGN-dominated LINERs are plotted as
open circles. Also displayed in
this figure is a starburst model (Rieke et al. 1993)
with burst duration of 1\,Myr (FHWM) 
(see figure caption for a detailed explanation). 
For the LINERs the Br$\gamma$ fluxes
are derived from the Pa$\beta$ fluxes using the theoretical value
given in Hummer \& Storey (1987), whereas the Br$\gamma$ equivalent
width is estimated from the Pa$\beta$ equivalent width and assuming
a color $J-K =1$ (which is almost independent of the stellar population).
This figure shows that both the [Fe\,{\sc ii}]$1.644\,\mu$m/Br$\gamma$ line
ratio and $K/{\rm Br}\gamma$ can be used to constrain the age of the burst
in those cases in which the LINER activity is dominated by
star formation. We find that the weak-[O\,{\sc i}] LINERs NGC~253, NGC~3367,
NGC~3504, NGC~4569 and NGC~4826 have burst ages (defined as the time
elapsed after the peak of star formation) between 8 and 11\,Myr,
whereas the classical LINERs NGC~404 and NGC~4736 seem to be in a
more advanced stage, with burst ages $> 12\,$Myr. The LINER/Wolf-Rayet
galaxy NGC~6764 has an age $\simeq 9-10\,$Myr. This model would only work 
if the burst duration is short compared with to the evolutionary 
lifetimes of massive stars as in the case of the 1\,Myr burst.

\begin{figure*}
\figurenum{7}
\plotfiddle{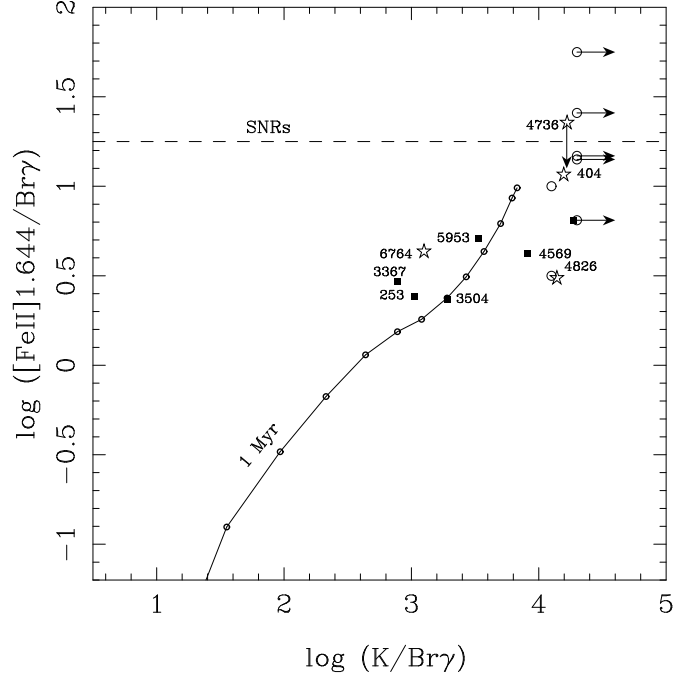}{425pt}{0}{60}{60}{-160}{165}
\vspace{-6.cm}
\caption{[Fe\,{\sc ii}]$1.644\,\mu$m/Br$\gamma$ vs.
$\log (K/{\rm Br}\gamma)$ diagram. The units for
$\log (K/{\rm Br}\gamma)$ are $\mu{\rm m}^{-1}$.
The solid line is the output of
a starburst model (Rieke et al. 1993) with the [Fe\,{\sc ii}]
calibration described
in Vanzi \& Rieke (1997) for a Gaussian burst of duration
1\,Myr (FWHM). For this curve the points are spaced
1\,Myr, and the first point on the curve (to the left) represents 2\,Myr
after the peak of star formation.  
The filled squares (our sample) and star-like symbols
(Calzetti 1997 and Larkin et al.  1998) are starburst-dominated
LINERs. The open circles are AGN-dominated LINERs. The
dashed line represents the average [Fe\,{\sc ii}]$1.644\,\mu$m/Br$\gamma$
line ratio for SNRs.}
\end{figure*}

The idea that some LINERs are powered by stars is not new;
for instance Filippenko \& Terlevich (1992) and Shields (1992)
proposed that the optical line ratios of LINERs can be explained in terms of
photoionization by hot stars ($ \geq 45,000\,$K) in a dense medium. However
more recently, Engelbracht et al. (1998) derived an upper limit of $37,000\,$K
on the stars exciting the emission lines in the weak-[O\,{\sc i}] LINER/H\,{\sc ii} galaxy NGC~253, based on different arguments. One of those arguments involves the observed value of the He\,{\sc i}($1.70\,\mu$m)/Br10 ($<0.25$) line ratio, which is predicted to be around 0.35 for hot stars ($T > 40,000\,$K).
For two galaxies in our sample, NGC~3367 and NGC~3504, we set the
same upper limit for the stellar temperature (Section~3.1).
The second argument used by Engelbracht et al. (1998)
is that hot stars are not needed to reproduce the optical line ratios of NGC~253. These authors assume that all the [Fe\,{\sc ii}] line emission is produced by supernovae (Vanzi \& Rieke 1997 and references therein), and use
photoionization models for metal-rich H\,{\sc ii} regions (Shields \&
Kennicutt 1995) to predict the line ratios for photoionized gas. Combining the
two excitation mechanisms produced a satisfactory fit to the overall spectrum
of the galaxy. In this fit, supernovae contribute a significant portion of
optical lines such as [O\,{\sc iii}]$\lambda5007$, [O\,{\sc i}]$\lambda6300$
and [S\,{\sc ii}]$\lambda\lambda6713,6731$.

\begin{figure*}
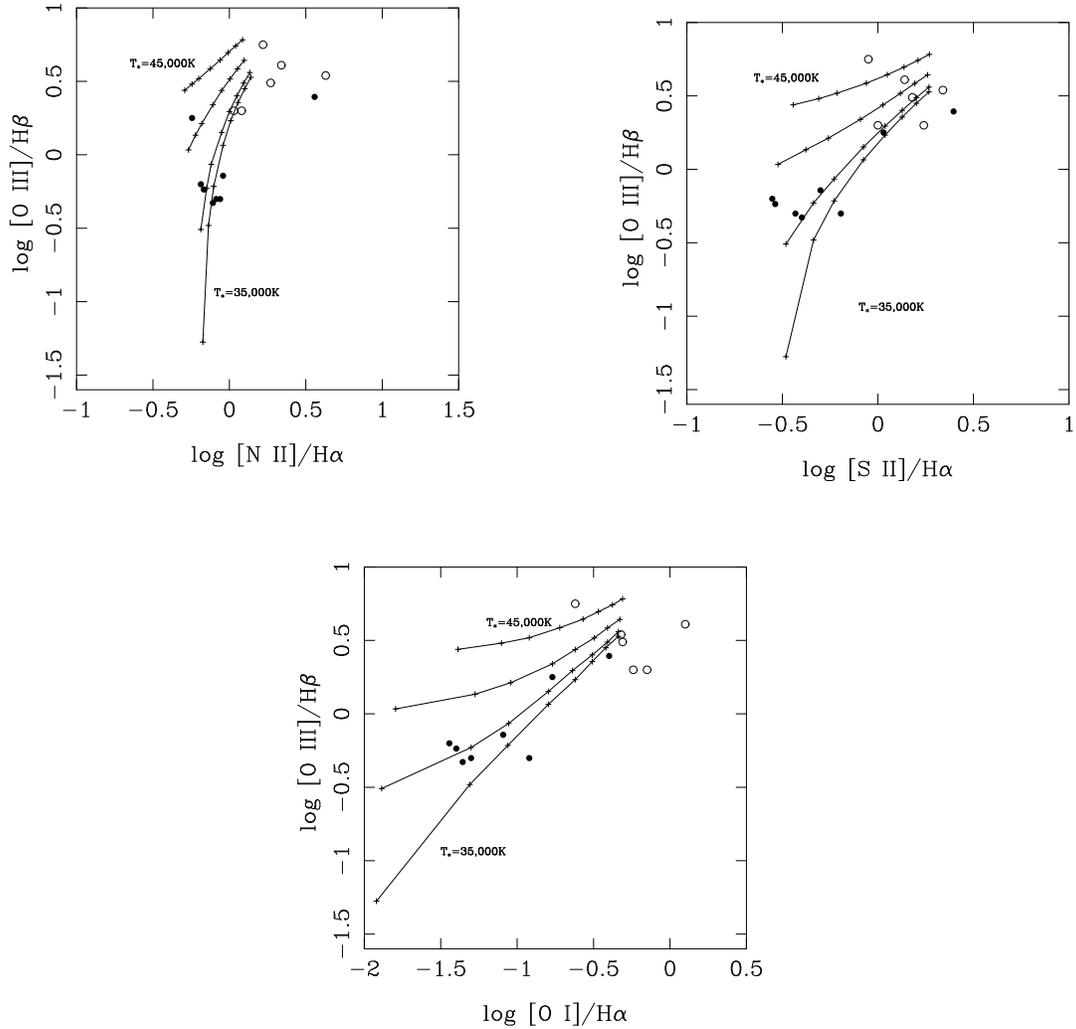

\figurenum{8}
\plotfiddle{figure8a.ps}{425pt}{0}{50}{50}{-230}{150}
\plotfiddle{figure8b.ps}{425pt}{0}{50}{50}{0}{580}
\plotfiddle{figure8c.ps}{425pt}{0}{50}{50}{-120}{810}
\vspace{-27cm}
\caption{Line ratio diagrams where the mixing curves of
H\,{\sc ii} region model with temperatures
$T=35,000, 38,000, 42,000$ and 45,000\,K  and SNR are plotted
as solid lines. The filled circles are the optical line ratios of
those starburst-dominated LINERs, whereas the open
circles are line ratios of the AGN-dominated LINERs. The symbols on
the curves indicate indicate [Fe\,{\sc ii}]$1.257\,\mu$m/Pa$\beta$
line ratios from left to right of 0. (that is, pure
H\,{\sc ii} region), 0.5, 1, 2, 3, 4, 5, and 6.}
\end{figure*}

We will take a similar approach to try and reproduce the line ratios of those
LINERs without evidence for an AGN. We determine the [Fe\,{\sc
ii}]$1.257\,\mu$m/Pa$\beta$ line ratios for the supernovae
together with the [O\,{\sc i}]$\lambda6300$ and
[S\,{\sc ii}]$\lambda\lambda$6713,6731
from the study of Lumsden \& Puxley (1995), which allows
us to obtain average values for the
[O\,{\sc i}]$\lambda 6300$/[Fe\,{\sc ii}]$1.257\,\mu$m
and [S\,{\sc ii}]/[Fe\,{\sc ii}]$1.257\,\mu$m line ratios. The data
for the [O\,{\sc iii}]$\lambda5007$ and [N\,{\sc ii}]$\lambda6583$ lines
are from Smith et al. (1993) and Blair \& Kirshner (1985).  All these 
lines are ratioed to [Fe\,{\sc ii}]$1.257\,\mu$m through hydrogen 
recombination lines, assuming Case B line ratios for the latter lines. 
To represent the H\,{\sc ii} region component we use the models presented in
Shields \& Kennicutt (1995). We choose models with solar metallicity,
including dust effects, and photoionized by a single star with
four different temperatures: $T = 35,000, 38,000, 42,000$ and $45,000\,$K
(calculated as described in Shields \& Kennicutt 1995 using CLOUDY 90.03).
For each galaxy, we use the [Fe\,{\sc ii}]$1.257\,\mu$m/Pa$\beta$ line
ratio to set the contribution of SNRs to the observed optical line
ratios. In Figure~8 we display mixing curves of an H\,{\sc ii}
region component with the different stellar temperatures and the
SNR component. The symbols on
the lines indicate [Fe\,{\sc ii}]$1.257\,\mu$m/Pa$\beta$
line ratios from left to right of 0. (that is, pure
H\,{\sc ii} region), 0.5, 1, 2, 3, 4, 5, and 6.
The starburst-dominated LINERs are  plotted as filled circles,
whereas the AGN-dominated LINERs are plotted as open circles.
From these diagrams (see also Table~7)
we observe that the mixed model with stellar
temperature $T=38,000\,$K provides a reasonable fit to almost all the
potentially starburst-dominated LINERs. 

\begin{deluxetable}{lccccc}
\tablefontsize{\footnotesize}
\tablewidth{18cm}
\tablecaption{Observed line ratios, and model predictions
($T_* = 38,000\,$K + SNR) for the starburst-dominated LINERs.}
\tablehead{\colhead{Galaxy} & \colhead{[O\,{\sc iii}]5007/H$\beta$} &
\colhead{[O\,{\sc i}]6300/H$\alpha$} &
\colhead{[N\,{\sc ii}]6583/H$\alpha$} &
\colhead{[S\,{\sc ii}]6713,6731/H$\alpha$} &
\colhead{[Fe\,{\sc ii}]/Pa$\beta$}}
\startdata
NGC~253  & 0.47 & 0.044 & 0.78 & 0.40 & $0.56\pm0.05$ \\
model    & $0.55\pm0.03$ & $0.053\pm0.003$ & $0.78\pm0.01$ & $0.47\pm0.01$ & \\
\hline
NGC~404  & $1.29-1.78$  & $0.17-0.23$ & $0.44-0.57$  & $0.94-1.03$ & 2.70 \\
model    & 1.73  & 0.21 & 1.03  & 1.03 \\
\hline
NGC~3367 & 0.50 & $0.031-0.050$ & $0.83-0.87$ & $0.28-0.37$  & $0.58\pm0.05$ \\
model    & $0.56\pm0.03$ & $0.055\pm0.003$ & $0.79\pm0.01$ & $0.48\pm0.01$ \\
\hline
NGC~3504 & $0.53-0.63$ & $0.023-0.036$ & $0.59-0.65$ & 0.28  & $0.54\pm0.05$ \\
model    & $0.53\pm0.03$ & $0.050\pm0.003$ & $0.78\pm0.01$ & $0.46\pm0.01$ \\
\hline
NGC~4569 & 0.72\tablenotemark{a}$-1.18$ & $0.062-0.081$ &
0.90 & $0.40-0.50$ & $1.50\pm 0.5$ \\
model    & $1.05\pm0.26$ & $0.12\pm0.035$ & $0.89\pm0.06$ & $0.72\pm0.13$ & \\
\hline
NGC~4736 & $1.47-$2.48\tablenotemark{a} & $0.24-0.40$ & $2.15-3.61$
& $1.39-2.49$ & 5.30 \\
model    & 3.17  & 0.41 & 2.35 & 1.98 & \\
\hline
NGC~4826 & $0.50-1.45$ & $0.073-0.12$ &  $0.82-1.27$ & $0.64-0.77$ & 0.70 \\
model    & 0.63 & 0.07 &  0.80 & 0.55 \\
\hline
NGC~6764 & 0.58 & $0.040-0.065$ & $0.68-0.75$ & $0.29-0.46$ & $1.0\pm 0.2$ \\
model    & $0.79\pm 0.10$  & $0.087\pm0.015$ & $0.83\pm0.02$ &
$0.59\pm 0.05$ \\
\enddata
\tablenotetext{a}{H$\beta$ flux derived from H$\alpha$, assuming
H$\alpha$/H$\beta = 2.70$.}
\tablecomments{References for the optical line ratios are as follows,
NGC~253 Armus, Heckman, \& Miley (1989), and Tadhunter et al. (1993);
NGC~404 Keel (1983), Ho et al. (1993, 1997a); NGC~4736 and NGC~4826 Keel
(1983) and Ho et al. (1997a); NGC~6764 Eckart et al. (1996).}
\end{deluxetable}

As an illustration of the calculations, we present in Table~7
the line ratios both observed and predicted by the model with
stellar temperature $T = 38,000\,$K for all the galaxies with evidence
for star-formation. Two basic assumptions are used in this
calculation. First we are taking average line ratios for SNRs, while an age
effect could be present, that is, more evolved SNRs show slightly different
average line ratios. Second, we are representing the H\,{\sc ii} region
component as photoionized by stars with a single temperature. This may
be true for the intermediate age starburst LINERs, but for the most evolved
starburst LINERs this may not be a good representation. For instance for
ages $> 5\,$Myr the [N\,{\sc ii}]$\lambda6583$/H$\alpha$,
[O\,{\sc i}]$\lambda6300$/H$\alpha$
and [S\,{\sc ii}]$\lambda\lambda$6713,6731/H$\alpha$ ratios slightly increase
as the starburst evolves,
whereas the [O\,{\sc iii}]$\lambda5007$/H$\beta$ ratio decreases rapidly
as the number of hot stars drops (see for example the models of
Garc\'{\i}a-Vargas, Bressan, \& D\'{\i}az 1995, and Stasi\'nska \&
Leitherer 1996). This last ratio is however easily maintained with the
presence of SNRs.

In all cases but NGC~6764 the observed line
ratios were measured from stellar continuum subtracted spectra (see references in the notes of the table). The range of observed line ratios corresponds to different references, and gives an idea of the uncertainties associated with these numbers. The errors associated with the line
ratios predicted by our H\,{\sc ii}-SNR models are simply the propagation
of the errors in the measured [Fe\,{\sc ii}]$1.257\,\mu$m/Pa$\beta$ line ratio. A 10\% error in the determination of the [Fe\,{\sc ii}]$1.257\,\mu$m/Pa$\beta$ line ratio would lead to errors in the model predictions of about 10\% in the
[O\,{\sc iii}]$\lambda5007$/H$\beta$ line ratio,
1\% in [N\,{\sc ii}]$\lambda$6583/H$\alpha$, 10\% in
[O\,{\sc i}]$\lambda 6300$/H$\alpha$
and 5\% in [S\,{\sc ii}]$\lambda\lambda$6713,6731/H$\alpha$.
The shock excitation produced by the SNRs accounts for
between 72\% and 100\% of the [O\,{\sc i}]$\lambda$6300 line,
between 25\% and 80\% of the [S\,{\sc ii}]$\lambda\lambda$6713,6731
lines, between 6\% and 44\% of the [N\,{\sc ii}]$\lambda$ 6583 line, and
between 42\% and 69\% of the [O\,{\sc iii}]$\lambda$5007 line  in the LINERs analyzed here.

From Table~7 we find that there is a very good agreement between the observed
and the predicted ratios for NGC~253, NGC~3367 and NGC~4569, and reasonably
good correspondence for NGC~404, NGC~3504, NGC~4736 and NGC~4826. NGC~6764 
shows
optical line ratios very similar to those of NGC~3504, however its
[Fe\,{\sc ii}]$1.257\,\mu$m/Pa$\beta$ line ratio is twice as large; one
reason for the discrepancy of the predicted line ratios could be the
difference in aperture size between the optical and the near-infrared line
ratios. The largest discrepancies seem to occur for the
[S\,{\sc ii}]$\lambda\lambda$6713,6731/H$\alpha$ line ratio,
the model always giving an overproduction of [S\,{\sc ii}]. However, the
average [S\,{\sc ii}]$\lambda\lambda$6713,6731/[Fe\,{\sc ii}]$1.257\,\mu$m ratio derived for SNRs is quite uncertain since this line ratio ranges from 2.87 to 10.52 (Lumsden \& Puxley 1995), making the predicted ratios
uncertain by a factor of two.

Despite the fact that our fitting technique is not very sophisticated,
it provides a very plausible fit to the line ratios in most of the
weak-[O\,{\sc i}]6300 and some classical LINERs. A separate set of 
calculations confirms that we can account for the line ratios in many of 
these galaxies with very hot (T $>$ 45,000K) stars as was pointed out by 
Filippenko \& Terlevich (1992). Thus, neither set of fits is unique for 
starburst-dominated LINERs, and the choice between them depends on other 
evidence. In this regard, the upper limits of 40,000K for the stellar 
temperature we have set for NGC~3504 and NGC~3367 would indicate that 
the new class of fit is to be preferred. Second, the high-stellar-temperature 
models have trouble reproducing cases with strong [O\,{\sc iii}] as well as 
low-ionization lines, such as those toward the upper right in Figure~8. 

Very hot stars appear to dominate starbursts for very short times. 
Therefore, we believe that the success of the lower temperature/shock 
excitation model is strong evidence that many LINERs are aging starbursts,
in which the SNRs play an increasingly important role as the starburst ages.
Not all the LINERs studied in this paper fall into either of these 
categories though. NGC~2639, NGC~1052, NGC~3031, NGC~3998, NGC~4579 
and NGC~7743 have optical line ratios which are not easily modeled 
with the composite H\,{\sc ii}
region -- SNR model. Although we do not rule out the possibility that
star formation is playing some role, it seems that in this group of objects
the low-luminosity AGN is dominant.

\section{Test of the LINER-Starburst Connection}

\subsection{General Behavior}

We can explore the LINER-Starburst connection through a simple parametarization of the behavior of a starburst galaxy. We take a typical mass-luminosity relation:

\begin{equation}
L(M) = K\, M^\alpha
\end{equation}

\noindent where $M$ is the stellar mass and $L$ its luminosity (and $K$
a constant). $\alpha$ is $\simeq 3$ for stars 
with $M> 10\,$M$_\odot$ (Vacca, Garmany, \& Shull 1996).
We approximate the main sequence lifetime of a star as,

\begin{equation}
T_{\rm ms} \propto \frac{M}{L}
\end{equation}

Now we assume a starburst in which all stars are formed simultaneously. It can
then be shown that the mass at the main sequence turnoff as a function
of the starburst age ($t$) is approximately:

\begin{equation}
M_{\rm ms}(t) = K_1\,t^{-\frac{1}{\alpha - 1}}
\end{equation}

\noindent where $K_1$ is a constant. We take the initial mass function IMF (for high mass stars) in the starburst to be (with upper mass cutoff $M_{\rm u}$):

\begin{equation}
\phi(M)\,{\rm d}M = C\,M^{\gamma}\,{\rm d}M,
\end{equation}

\noindent where $C$ is a constant.
With these approximations, the supernova rate (SNr) can be computed as,

\begin{equation}
{\rm (SNr)} = \phi(M)\,\frac{{\rm d}M_{\rm ms}}{{\rm d}t}
\propto t^{\frac{-\gamma - \alpha}{\alpha - 1}}
\end{equation}

\noindent Using a similar approach, the starburst luminosity is:

\begin{eqnarray}
L_{\rm sb}  = \int_{M_{\rm l}}^{M_{\rm u}}\phi(M)\,L(M)\,{\rm d}M
\simeq\\
\int_{M_{\rm l}}^{M_{\rm ms}}\phi(M)\,L(M)\,{\rm d}M
\propto M_{\rm ms}^{\gamma + \alpha + 1} \propto
t^{-\frac{\gamma + \alpha + 1}{\alpha - 1}}
\end{eqnarray}

\noindent We have made the approximation that all the luminosity emerges from
stars near the main sequence turnoff ($M_{\rm ms}$)
in simplifying the integral. Additional
luminosity is produced by stars in post-main-sequence evolution, but
it would correct the expression by only a small factor, and since
these stages are predominantly at much lower stellar temperature than
the main sequence ones, much of this luminosity may escape the starburst
region. Thus, from equations~[5] and [6] we get that
the starburst luminosity to supernova rate ratio does not depend upon
the IMF slope, and using the nominal value of $\alpha$ goes as,

\begin{equation}
\frac{L_{\rm sb}}{{\rm (SNr)}}
\propto t^{\frac{-1}{\alpha - 1}} \propto t^{-0.5}
\end{equation}

\noindent That is, so long as the main sequence turnoff is high enough that virtually all evolving stars end their lives as supernovae (nominally corresponding to a turnoff mass near 8\,M$_\odot$), then the proportionality between luminosity and supernova rate should change only slightly as the starburst ages.

However, because starburst galaxies are usually identified by their
emission lines, it would be more appropriate to use the ionizing luminosity to
represent the starburst phase. The ionizing luminosity can be
expressed as $L_{\rm ion} \propto M^\beta$, and in this case, equation~[7] becomes,

\begin{equation}
\frac{L_{\rm ion}}{{\rm (SNr)}}
\propto t^{\frac{\alpha -\beta - 1}{\alpha - 1}}
\end{equation}

\noindent The value of $\beta$ is approximately 7 for dwarf 
stars with effective temperatures between 32,000\,K and 
40,000\,K  (Vacca et al. 1996), so the time 
dependence is: 

\begin{equation}
\frac{L_{\rm ion}}{{\rm (SNr)}}
\propto t^{-2.5}
\end{equation}

\noindent The analytical prediction given in Equation~(9) is in good agreement
with the model prediction (Figure~7) in which the line ratio 
[Fe\,{\sc ii}]$1.644\,\mu$m/Br$\gamma$ drops by more than a 
factor of a hundred between 1 and 10\,Myr.

Thus, the nature of the emission spectrum of a galaxy will change 
rather abruptly from H\,{\sc ii}-dominated to shock-dominated (Equation~9). 
This transition will occur while there is still a substantial luminosity 
from the starburst (Equation~7). We therefore predict that there will be 
two general classes of luminous starburst-powered galaxy, dominated 
respectively by hot stellar and shock excitation. Intermediate cases 
will arise largely because extended periods of star formation can cause 
parts of the starburst to be in a young phase while other parts have 
evolved into the shock-excited phase. 

\subsection{The relative numbers of starbursts and starburst-dominated 
LINERs}
To test these predictions, we need to determine when the supernova rate is high
enough (compared with the luminosity of the starburst) that we will start
detecting LINER characteristics. Pure starbursts could be defined such that 
the ratio of ionizing radiation to supernova shock-heating is greater than 1.
From Figure~7, for a burst of 1\,Myr duration (FWHM), the transition occurs after 
about 6\,Myr, as defined by when the ratio of the predominantly shock excited 
[Fe\,{\sc ii}] to the predominantly photo-ionized Br$\gamma$ lines exceeds one. In 
fitting starbursts with evolutionary models, it is generally found that strong 
star formation occurs over a period of $5-10\,$Myr (e.g., Rieke et al. 1993; 
Genzel et al. 1995; Engelbracht et al. 1996, 1998; B\"oker et al. 
1997; Vanzi et al. 1998). The lifetime of the LINER phase could be 
defined when the supernova rate drops by a factor of 10 from its peak value (after 
30 to 40\,Myr). We choose 30\,Myr as the  maximum age at which we would be able to detect
H$\alpha$ in emission, for at older ages the equivalent width of this line would 
be $<0.5$\AA \ (e.g., Leitherer \& Heckman 1995; Leitherer et al. 1999).
The predicted behavior of a starburst is then about 14 Myr of photo-ionization 
domination, followed by about 30 Myr of shock domination. The expected relative 
number between (young) starburst and starburst-dominated LINERs (aging
starbursts) is simply given by the ratio between the durations of 
each phase, that is, roughly 1:2. 

The distance-limited Palomar spectroscopic survey of nearby galaxies 
(Ho et al. 1997a,c) can be used to test this prediction. From the sample of 
486 galaxies, 206 galaxies are classified as H\,{\sc ii} nuclei. Approximately 
100 galaxies classified as H\,{\sc ii} nuclei show equivalent widths of 
H$\alpha$ of less than 10\,\AA. Among these galaxies we find that approximately 
50 of them show [O\,{\sc i}]/H$\alpha > 0.040$,  
whereas from figure~7 in Ho et al. (1997a) the arbitrary line for 
transition objects has been drawn around [O\,{\sc i}]/H$\alpha = 0.080$.  
From Figure~8b we can see that objects with [O\,{\sc i}]/H$\alpha > 0.040$
have already a significant contribution from SNRs, and could be considered 
(arbitrarily) as evolved starbursts. In addition, for those objects with
small values of EW(H$\alpha$) the [O\,{\sc i}] line is weak, and the 
measurement errors are of the order of 20\% or 30\% (Ho et al. 1997a). 
Because of these ambiguities, we exclude the galaxies with very faint 
H$\alpha$ from the sample, leaving roughly 100 starburst galaxies strong
 enough that we might expect to detect their counterparts as LINERs. 94 
sample members are classified as LINERs and 65 galaxies are classified as 
transition objects (i.e., in Ho et al. 1997c definition these are composite 
LINER/H\,{\sc ii} nuclei). About 20\% of all AGN/LINER identifications show 
broad H$\alpha$ and are presumably powered by a true AGN. We have subtracted 
this percentage from the
the LINER category to isolate the candidate starburst-dominated
LINERs. We also exclude the LINER/Seyferts from the LINER category.
We find that the relative number of starbursts to starburst-dominated LINERs 
in the Ho et al. (1997a) sample is about 1:1.4 which is in satisfactory 
agreement with the prediction, given the necessarily rough nature of the estimates.

\subsection{Morphological Types}

Starbursts appear preferentially in late type spiral galaxies, 
while LINER nuclei are found mostly in early type galaxies, including 
ellipticals and S0s. We have examined this trend in further detail using 
the sample with very high quality spectra described by Ho et al. (1997b). 
We consider only the galaxies with pure LINER spectra. If galaxies with 
indications of an AGN are excluded (broad wings on H$\alpha$, strong radio 
or X-ray emission), then we find only one elliptical remains within the 
sample. The morphological type distribution is centered at about Sa and 
is not significantly different from that for LINER/Transition galaxies. 

Thus, the galaxies that plausibly contain aging starbursts are mostly of 
spiral type, but are earlier than the bona fide starburst galaxies. An 
interesting possibility is that the consequences of a strong nuclear starburst 
are to modify the appearance of a galaxy so that it tends to be classified 
as of an earlier type than before the burst. This change could result from 
the exhaustion of interstellar gas in the center and the presence of a 
luminous population of intermediate population stars with the bulge that 
increase its prominence. An example is provided by NGC 4736, which is a
 LINER of type Sab. However, a number of lines of evidence suggest that 
its circumnuclear stellar population is dominated by the products of an 
aging starburst (e.g., Walker, Lebofsky, \& Rieke 1988; Taniguchi et al. 
1996). The tendency of a number early type LINERs [NGC 404 (S0), NGC 4569 
(Sab), NGC 5055 (Sbc), and NGC 6500 (Sab)] to have UV spectra dominated by 
hot, massive stars (T $>$ 30,900K) is further evidence that there is a 
connection between starbursts and early type LINERs (Maoz et al. 1998). 

\section{CONCLUSIONS}

We have presented $J$-band ($1.15-1.35\,\mu$m) spectroscopy of a sample 
of nine galaxies showing some degree of LINER activity (classical
LINERs, weak-[O\,{\sc i}] LINERs and transition objects),
together with $H$  spectroscopy
for some of them. All the LINERs show bright [Fe\,{\sc ii}]$1.2567\,\mu$m 
emission, whereas the Pa$\beta$ line is only detected in emission in three
LINERs (before stellar continuum subtraction). Since
the $J$-band spectra of most LINERs in this work are
dominated by a strong stellar continuum, we perform a careful subtraction 
of this continuum to measure more accurate 
[Fe\,{\sc ii}]$1.2567\,\mu$m/Pa$\beta$ ratios. 

We find that the optical and near-infrared emission lines of a
significant percentage of LINERs (those with no evidence for an AGN) 
can be explained by the presence of an evolved starburst. We model 
the optical line ratios of these LINERs with a metal-rich H\,{\sc ii} region
model with $T_{\rm *} = 38,000\,$K, plus a SNR component.
The proportions of the two components are set by the
[Fe\,{\sc ii}]$1.2567\,\mu{\rm m/Pa}\beta$ line
ratio, since the [Fe\,{\sc ii}] line is predominantly excited by
SNR-driven shocks and the line Pa$\beta$ tracks H\,{\sc ii} regions
excited by massive young stars.
The proposed LINER-starburst connection is tested by predicting
the time dependence
of the ionizing luminosity of the starburst to the supernova rate
$L_{\rm ion}$/(SNr). We predict the relative number of
starbursts to starburst-dominated LINERs (aging starbursts).
The agreement with the observations is relatively good, supporting our
hypothesis that many LINERs are aging starbursts.

$\,$

$\,$

We thank the anonymous referee for providing comments which helped to
improve the paper. We are grateful to Ana Maria Biscaya and Alice Quillen for assisting us with some of the observations presented in this paper.

During the course of this work AA-H was supported by the National Aeronautics
and Space Administration on grant NAG 5-3042 through the University of
Arizona. The work was also partially supported by the National Science
Foundation under grant AST-95-29190. This research has made use of the
NASA/IPAC Extragalactic Database (NED) which is operated by the Jet Propulsion
Laboratory, California Institute of Technology, under contract with the
National Aeronautics and Space Administration.

\bibliographystyle{mn}

\appendix

In this appendix we discuss the properties of the sample of LINERs presented
in this paper together with those additional LINERs for which there is
significant evidence for the starburst activity to be related with
the LINER activity.

INDIVIDUAL OBJECTS

{\it NGC~253.---} The properties of this galaxy are extensively discussed in
Engelbracht et al. (1998). Its optical line ratios place it as
a transition object between pure starburst and weak-[O\,{\sc i}] LINERs.
As shown in Engelbracht et al. (1998) and here, the optical line ratios
of this galaxy are consistent with photoionization by stars with
$T = 38,000\,$K and some contribution from SNRs. Therefore the LINER
characteristics of this galaxy are entirely explained by the
presence of a starburst.

{\it NGC~404.---} The optical line ratios in this galaxy (Ho et al. 1993)
meet Heckman (1980) definition for LINERs. No broad component of
H$\alpha$ is detected (Ho et al. 1997b). The location in the optical
ratio diagnostic diagrams places this galaxy in the
transition object region.
Evidence for star-formation in this galaxy comes from both
the young underlying
stellar population and the circumnuclear H\,{\sc ii} region-type emission
in its optical spectrum (Ho et al. 1993, 1995).  In addition the UV
ionizing source in this
galaxy is a star cluster slightly older than the one in NGC~4569 (Maoz
et al. 1998), as also suggested by the higher SNR contribution than
in NGC~4569 needed to
reproduce the [Fe\,{\sc ii}]$1.257\,\mu$m/Pa$\beta$ line
ratio and optical line ratios (Table~7 and Figures~7 and 8). NGC~404 falls
in the starburst-dominated LINER category.

{\it NGC~1052.---} This galaxy is the ``prototypical'' LINER. NGC~1052 is
now known to show broad Balmer lines in polarized light (Barth et al. 1999),
which makes it another AGN-dominated LINER. The properties
of this LINER together with the infrared
spectroscopy were discussed in detail in Alonso-Herrero et
al. (1997).

{\it NGC~2639.---} The optical line ratios (Ho et al. 1993)
of NGC~2639 are very close to Heckman's definition of LINER.
The presence of a broad H$\alpha$ component in this
galaxy has long been known (Keel 1983; Huchra et al. 1982; Ho et al.
1997a). The [Fe\,{\sc ii}]$1.257\,\mu$m line emission is very strong
with Pa$\beta$ being observed probably in absorption. It shows the
highest [Fe\,{\sc ii}]$1.257\,\mu$m/Pa$\beta$ line
ratio in our sample, in good accordance with the high
[O\,{\sc i}]$\lambda 6300$/H$\alpha$ line  ratio. This galaxy is a
good example of AGN-dominated LINER. Note that this galaxy
contradicts Larkin et al. (1998) claim that LINERs with high
[Fe\,{\sc ii}]$1.257\,\mu$m/Pa$\beta$ ratio are starburst-dominated
LINERs.

{\it NGC~3031 (M81).---} The LINER M81 is one of the best candidates for
low-luminosity AGN (i.e., AGN-dominated LINER in our classification).
Its properties in the UV and optical have been recently studied in
great detail by Ho et al. (1996). Our $J$-band spectrum shows
[Fe\,{\sc ii}]$1.257\,\mu$m emission embedded in a very strong
underlying stellar continuum. The [Fe\,{\sc ii}] emission must be
very localized at the center of the galaxy, since the off-nucleus
spectrum (see Figure~2) shows no emission from this line. The Pa$\beta$
line is probably seen in absorption. Due to the radial velocity of this
galaxy, a small error in the correction for the Pa$\beta$ emission
introduced by the division by the standard star can affect the detection of
a very faint Pa$\beta$ emission from M81.

{\it NGC~3367.---} NGC~3367 is of special interest because
its optical line ratios do not correspond to those of an H\,{\sc ii}
region photoionized by hot stars, or  a Seyfert 2 or a LINER (Dekker et al.
1988). The presence of a broad component of H$\alpha$ is not clear, due
to important asymmetries (Ho et al. 1997b). In the optical it shows evidence
for the presence of Wolf-Rayet emission features (Ho et al. 1995).
This galaxy shows one of the lowest [Fe\,{\sc ii}]$1.257\,\mu$m/Pa$\beta$
line ratios in our sample, giving indication for youth of the star-formation
process. The upper limit to the He\,{\sc i}/Br10 line ratio
sets an upper limit to the stellar temperature ($<40,000\,$K) which is
in good agreement with the stellar temperature needed in the
H\,{\sc ii}/SNR model to reproduce its optical line ratios.
NGC~3367 is clearly a starburst-dominated transition object.

{\it NGC~3504.---} This galaxy has been classified as a
weak-[O\,{\sc i}] LINER (Ho et al. 1993) and as a H\,{\sc ii} region
(Ho et al. 1997a). In Alonso-Herrero et al. (1997) we presented $H$ and
$K$-band spectroscopy of this galaxy and concluded that the LINER
characteristics of NGC~3504 could be easily explained with the presence of
a recent starburst. The age of the burst, based on the
[Fe\,{\sc ii}]$1.257\,\mu$m/Pa$\beta$ line ratio would be very
similar to that of NGC~253. From Table~7 and Figure~8 it is clear that the
optical line ratios of this galaxy are well reproduced with an H\,{\sc ii}
region photoionized with stars with temperature $T = 38,000\,$K without
the presence of very hot stars.

{\it NGC~3998.---} The optical line ratios (Ho et al. 1993)
of NGC~3998 meet the LINER definition. The detection of broad
H$\alpha$ (see Ho et al. 1997b and references therein) makes NGC~3998 another
candidate for AGN-dominated LINER.
This galaxy is in common with Larkin et al. (1998) study of LINERs. Their
measured [Fe\,{\sc ii}]$1.257\,\mu$m/Pa$\beta$ line ratio is
lower than ours. Comparing our spectrum with theirs, our
detection of the [Fe\,{\sc ii}]$1.257\,\mu$m line has a  higher
signal to noise. Larkin et al. (1998)
claimed that the [Fe\,{\sc ii}]$1.257\,\mu$m/Pa$\beta$ ratio was
significantly lower in comparison with the [O\,{\sc i}]$\lambda 6300$/H$\alpha$
ratio. However with our
[Fe\,{\sc ii}]$1.257\,\mu$m/Pa$\beta$ line ratio measured from the
stellar continuum subtracted spectrum, the galaxy fits perfectly in
the {\it correlation} (see Figure~6). NGC~3998 is another
example of AGN-dominated LINER.

{\it NGC~4569 (M90).---} The optical line ratios of this galaxy given
in Stauffer
(1982) comply with Heckman definition of LINERs, however, it is classified
as a transition object in Ho et al. (1997a) based on both its location in the
diagnostic diagrams and the non detection of a broad component of H$\alpha$
(Ho et al. 1997b). The {\it HST}/UV spectrum of this galaxy (as in NGC~404)
shows absorption line signatures indicative of a continuum dominated
by light from massive stars (Maoz et al. 1998), and the {\it HST}/UV imaging
shows extended emission (Barth et al. 1998). Keel (1996) in a  detailed
study of the UV and optical properties reaches the conclusion that a
a population of young stars is responsible for the properties of
NGC~4569, although an extra component associated with
either an AGN or a population of older stars is still necessary.
Our $J$ and $H$-band
spectra show that the emission is strongly dominated by the underlying
stellar continuum. The Pa$\beta$ line is clearly detected in emission.
All this evidence leads to a classification of starburst-dominated
LINER.

{\it NGC~4579.---} The optical line ratios of this galaxy (Gonz\'alez-Delgado
\& P\'erez 1996b) satisfy only one of the criteria in Heckman's
definition ([O\,{\sc ii}]$\lambda3727$/[O\,{\sc iii}]$\lambda5007 = 1.53$).
However the position of the line ratios in the
diagnostic diagrams is in
the overlapping region of LINERs and Seyferts which lead Ho et al. (1997a)
to classify this galaxy as a Seyfert 1.9 or LINER 1.9. The AGN-dominated nature
of this LINER is
confirmed by the detection of broad wings in the H$\alpha$ line
(Filippenko \& Sargent 1985; Ho et al. 1997b) along with broad lines
in the UV (Maoz et al. 1998). NGC~4579 is the only galaxy in our sample with
a FWHM of the [Fe\,{\sc ii}]$1.257\,\mu$m line well above the instrument
resolution. The $J$-band spectrum does not show Pa$\beta$ in emission.

{\it NGC~4736 (M94).---}
This {\it true} LINER, according to Heckman (1980) definition,
is an interesting example that suggests a combination of
excitation mechanisms. The stellar continuum subtracted optical
spectrum shows characteristics of a transition object (Ho et al. 1995), or
type-2 LINER (Ho et al. 1997a), rather
than a classical LINER.  It does not show a broad component of
H$\alpha$ (Ho et al. 1997b).
This LINER has strong Balmer optical and infrared absorptions (e.g.,
Larkin et al. 1998) and other
characteristics that make it the prototype of a late-phase starburst
(e.g., Walker, Lebofsky, \& Rieke 1988; Taniguchi et al. 1996), but also
has a compact X-ray source possible indicative of a weak AGN (Cui, Feldkuhn,
\& Braun 1997). The high value of the [Fe\,{\sc ii}]$1.257\,\mu$m/Pa$\beta$
line ratio (Larkin et al. 1998) confirms the presence of an evolved
starburst in which the SNRs have a dominant contribution.

{\it NGC~4826 (M64).---} This galaxy satisfies the LINER definition of
Heckman (1980), however it is
located in the region between transition objects and LINERs in Ho et al.
(1997a) diagrams.
No broad H$\alpha$ component is detected (Ho et al. 1997b).
The range of optical line ratios measured by Keel (1983) and Ho et al.
(1997a) are well reproduced with our model of H\,{\sc ii} region and
SNRs. This fact seems to indicate that NGC~4826 is a starburst-dominated
LINER. In addition, this galaxy shows a chain of H\,{\sc ii} regions
near the center and evidence for shock-excited gas some 4 arcseconds
off the nucleus (identified as a SNR by Rix et al. 1995).

{\it NGC~5953.---} This is an Sa pec galaxy interacting with the
LINER/starburst
galaxy NGC~5954. The determination of the activity class of the nucleus
of this galaxy has been somehow problematic.
It has been classified as a LINER by Veilleux et al. (1995), and as
a Seyfert 2 by Gonz\'alez-Delgado \& P\'erez (1996a). Note that some of optical
line ratios reported for this galaxy in these two works differ by
a factor of two. Using the reddening corrected optical line ratios given
in  Keel et al. (1985) and Veilleux et al. (1995), and the diagnostic
diagrams, we find that
NGC~5953 is located in the region occupied by both LINERs and transition
objects. It shows strong circumnuclear star-formation
activity as inferred from both 2D optical spectroscopic data
(Gonz\'alez-Delgado \& P\'erez 1996a; V\'eron,
Goncalves \& V\'eron-Cetty 1997) and {\it HST}/UV
images (Colina et al. 1997). Our $J$-band spectrum  shows
strong [Fe\,{\sc ii}] and Pa$\beta$ emission lines. The ratio
[Fe\,{\sc ii}]$1.2567\,\mu$m/Pa$\beta$ decreases for the large
aperture (Table~5) which is an indication for an increased
star-formation activity in the circumnuclear regions as seen
in the optical. $H$ and $K$-band
spectroscopy of the companion NGC~5954 is presented in Vanzi,
Alonso-Herrero, \& Rieke (1998).

{\it NGC~6764.---} This SBb galaxy is classified as a classical LINER galaxy
(i.e., satisfies Heckman's definition). It
shows Wolf-Rayet star features in its optical spectrum (Eckart et al 1996
and references therein), which makes this galaxy one
of the most significant examples of
starburst dominated classical LINER. The analysis of the starburst
properties of this galaxy (Eckart et al. 1996)
reveals that this galaxy is undergoing a burst of star-formation
with a time scale of $10^7\,$yr consistent with the age
derived from the [Fe\,{\sc ii}]$1.257\,\mu$m/Pa$\beta$
line ratio and EW of Br$\gamma$ (values taken from
Calzetti 1997) in Figure~7.

{\it NGC~7743.---} The optical line ratio [O\,{\sc
i}]$\lambda$6300/H$\alpha$ of NGC~7743 is typical of
weak-[O\,{\sc i}] LINERs. However,
when plotted in the diagnostic diagrams
(Ho et al. 1994, 1997a), it lies close to
the LINER/Seyfert 2 region. In the UV NGC~7743 is spatially extended
(Barth et al. 1998). Its optical line ratios cannot be fitted with our
H\,{\sc ii}-SNR model.

\end{document}